\documentclass[aps,a4paper, preprint, superscriptaddress,preprintnumbers,floatfix,nofootinbib,amsmath,amssymb]{revtex4}
\usepackage{url}
\usepackage{hyperref}
\usepackage{cleveref}

\usepackage{amstext,amssymb}
\usepackage[section]{placeins}
\usepackage{amsmath}
\usepackage{graphicx}
\usepackage{xspace}
\usepackage{color}
\usepackage{xcolor}
\usepackage{float}
\usepackage{comment}
\usepackage{amstext,amssymb}
\usepackage{amsmath}
\usepackage{graphicx}
\usepackage[section]{placeins}
\usepackage{xspace}
\usepackage{color}
\usepackage{units}
\usepackage{array}
\usepackage[section]{placeins}
\usepackage{amsmath,bm}
\usepackage{amssymb}
\usepackage{gensymb}
\usepackage{soul}
\usepackage{textcomp}
\usepackage{slashed} 
\usepackage{multirow}
\usepackage{hyperref}
\usepackage{mathrsfs}
\usepackage{enumitem}
\usepackage{graphicx}
\usepackage{color}
\usepackage{comment}
\usepackage{epstopdf}
\usepackage{float}
\usepackage{units}
\usepackage{feynmp}
\usepackage{slashed}
\usepackage{amsmath,bm}
\DeclareUnicodeCharacter{2212}{-}
\usepackage{slashed} 
\usepackage{multirow}
\newcommand{\be}{\begin{equation}}
\newcommand{\ee}{\end{equation}}
\newcommand{\bea}{\begin{eqnarray}}
\newcommand{\eea}{\end{eqnarray}}
\newcommand{\nn}{\nonumber}

\def\s1{\hat s}

\usepackage{slashed}

\newcommand{\nua}[1]{\ensuremath{\rlap{\kern-2.5pt\ensuremath{\overset{\scriptscriptstyle(-)}{\phantom{\nu}}}}{\ensuremath{{\nu}_{#1}}}}\xspace}

\begin{document}
\title{ Type III seesaw under $A_4$ modular symmetry with leptogenesis}
\author{Priya Mishra }
\email{mishpriya99@gmail.com}
\affiliation{School of Physics,  University of Hyderabad, Hyderabad - 500046,  India}
\author{Mitesh Kumar Behera}
\email{miteshbehera1304@gmail.com}
\affiliation{School of Physics,  University of Hyderabad, Hyderabad - 500046,  India}
\author{Papia Panda}
\email{ppapia93@gmail.com}
\affiliation{School of Physics,  University of Hyderabad, Hyderabad - 500046,  India}
\author{ Rukmani Mohanta}
\email{rmsp@uohyd.ac.in}
\affiliation{School of Physics,  University of Hyderabad, Hyderabad - 500046,  India}

\begin{abstract}
We make an attempt to study neutrino phenomenology in the framework of type-III seesaw by considering $A_4$ modular symmetry in the  super-symmetric context. In addition, we have included a local $U(1)_{B-L}$ symmetry which eventually helps us to avoid certain unwanted terms in the superpotential. Hitherto, the seesaw being type-III, it involves three fermion triplet superfields $\Sigma_R$, along with which, we have included a singlet weighton field $(\rho)$. In here, modular symmetry plays a crucial role by avoiding the usage of excess flavon (weighton) fields.  Also, the Yukawa couplings acquire modular forms which are expressed in terms of Dedekind eta function $\eta(\tau)$. However, for numerical analysis we use $q$ expansion expressions of these couplings. Therefore, the model discussed here is triumphant enough to accommodate the observed  neutrino oscillation data and also successfully explains observed baryon asymmetry of the universe through leptogenesis.
\end{abstract}

\maketitle
\flushbottom

\section{INTRODUCTION}
\label{sec:intro}
Decades ago when standard model (SM) was built it seemed impeccable, but its foundation was again questioned when some unresolved puzzles came into existence. To name a few,  it does not provide any satisfactory explanation to the tininess of neutrino mass \cite{Ma:1998dn}, neutrino oscillation,  strong CP problem, matter-antimatter asymmetry, the nature of dark matter and dark energy, etc. To resolve the issue regarding smallness of neutrino masses within the context of  SM, Weinberg operator \citep{Weinberg:1979sa,Abada:2007ux} helps to an extent. However, to demonstrate other phenomena, we need to go beyond standard model (BSM), so introducing right handed (RH) neutrinos becomes a necessity. This becomes the basis of  canonical seesaw mechanism, i.e., as soon as these RH neutrinos come into picture, they allow Dirac mass terms for neutrinos. In this regard, type-I seesaw \citep{Minkowski:1977sc,Yanagida:1979as,Glashow:1979nm,Mohapatra:1979ia} is the simplest one,  which includes singlet heavy ($\simeq 10^{14}$ GeV) RH neutrinos and brings down the mass scale of active neutrinos to $0.1$ eV range, as observed from experimental data. Also, there exists other variants of seesaw i.e., type-II \cite{Mohapatra:1980yp,Antusch:2004xy,Gu:2006wj,Arhrib:2011uy,Ghosh:2017pxl} which incorporates scalar triplets, type-III \cite{Foot:1988aq,Liao:2009nq} involving fermion triplets, linear seesaw \cite{Ma:2009du,Wang:2015saa,Borah:2018nvu,CarcamoHernandez:2020pnh,Sruthilaya:2017mzt} and inverse seesaw \cite{Hirsch:2009mx,Gu:2010xc,Das:2012ze,Arganda:2014dta,Dias:2012xp,Dev:2012sg,Dias:2011sq,Bazzocchi:2010dt} which are modified  type-I seesaw. In this work, we intend to study the case of type-III seesaw  in the context of discrete $A_4$ modular symmetry as it has not been studied earlier in this framework. In general, it is presumed that type III seesaw is more complicated compared to the canonical type I seesaw due to the involvement of triplet fermions. However, it has been shown in Refs. \cite{Barr:2003nn, Albright:2003xb} that, in some cases, e.g., realistic $SO(10)$ model, type III seesaw  may have less difficulty in reproducing realistic neutrino masses and mixings than the conventional type-I seesaw. Therefore, in this work we would like to investigate the implications of $A_4$ modular symmetry   in the context of type III seesaw for describing the observed neutrino oscillation data.


It is interesting to notice that many non-abelian discrete symmetries i.e., $S_3$ \citep{Ma:2004zd,Kubo:2003iw,Pakvasa:1977in,Ma:2014qra}, $A_4$ \citep{Ma:2001dn,Babu:2002dz,Altarelli:2005yp,Ma:2004zv}, $S_4$ \citep{Ma:2005pd,Krishnan:2012me,Grimus:2009pg} etc. and continuous symmetries like $U(1)_{B-L}$ \citep{Mishra:2020fhy,Ma:2015raa,Singirala:2017see,Singirala:2017cch,Nomura:2017jxb,Nomura:2017vzp,Nomura:2017kih}, $U(1)_{L_e-L_\mu}$ \citep{Behera:2021nuk,Foot:1990mn,Panda:2022kbn,He:1991qd}, $U(1)_{H}$ \citep{Nardi:2000px,Ibanez:1994ig,Binetruy:1994ru,Nir:1995bu} etc. come to our rescue to develop the model and generate neutrino mass matrix, which gives results in-accordance with experimental data. Implementation of the discrete non-abelian symmetries demand the usage of excess flavon fields. These flavon insertions make the Yukawa interaction terms non-renormalizable and bring down the predictability of the model. Therefore, a clever approach of modular symmetry \citep{Ferrara:1989bc,Ferrara:1989qb,Leontaris:1997vw,feruglio2019neutrino,King:2020qaj} is introduced to breach the scenario of flavon fields. In here, the approach involves discrete symmetry group because they are isomorphic to finite modular groups, for example, $\Gamma_2 \simeq S_3$ \citep{Okada:2019xqk,Kobayashi:2018vbk,Kobayashi:2018wkl,Kobayashi:2019rzp}, $\Gamma_3 \simeq A_4$ \citep{Behera:2020lpd,Nomura:2022boj,Kashav:2022kpk,Kashav:2021zir,Behera:2020sfe,Asaka:2020tmo,Abbas:2020vuy,Okada:2020dmb,Altarelli:2005yx}, $\Gamma_4 \simeq S_4$ \citep{Kobayashi:2019xvz,Wang:2019ovr,Okada:2019lzv,King:2019vhv,Kobayashi:2019mna,Novichkov:2018ovf,Penedo:2018nmg}, $\Gamma_5 \simeq A_5$ \cite{Criado:2019tzk,Novichkov:2018nkm,Ding:2019xna},  $\Gamma^\prime_5 \simeq A^\prime_5$ \citep{Behera:2021eut,Behera:2022wco,Yao:2020zml,Wang:2020lxk} etc. We make an attempt to use $A_4$ modular symmetry, which is isomorphic to $\Gamma_3$. The alluring feature of modular symmetry is that, it transforms Yukawa couplings i.e., it makes them modular in nature. Therefore, the flexibility to fine tune the Yukawa couplings is lost and now it is governed by the modulus $\tau$. The involvement of modulus $\tau$ is seen in the expression of Dedekind eta function, as shown in eqn.(\ref{dedekind}), and further the acquisition of VEV by it, helps in the symmetry breaking of the $A_4$ group. As, $N=3$ for $\Gamma_3 \simeq A_4$ is finite, hence, they can be constructed using $k=1$ being the lowest weight. The dimension of $\Gamma_3$ being $2k+1$ (see appendix D of \cite{feruglio2019neutrino}) yielding three linearly independent $Y_i(\tau)$ shown in eqns. (\ref{mod_y1} $-$ \ref{mod_y3}). These Yukawa couplings are utilised in curating the neutrino mass matrices after applying the $A_4$ product rules and are implicitly governed by the range of modulus $\tau$, as will become more clear while performing the analysis numerically. 
Further, we are able to explain the baryon asymmetry of the universe through leptogenesis \citep{Davidson:2008bu,Fukugita:1986hr}, because of presence of heavy RH neutrino, which yields the  order of lepton asymmetry to be $\sim 10^{-10}$.

This work is organised as follows. In Sec.\ref{sec:2}, we discuss the model framework containing particles contributing towards expressing the superpotential for type-III seesaw and its associated mass matrices. Further, in Sec.\ref{sec:numerical}, we perform the numerical analysis where a common parameter space along with best-fit data set are extracted using chi-square minimization technique using the data of all the phenomena discussed in our model.
Additionally, Sec.\ref{sec:lepto} sheds light on lepton asymmetry generated through leptogenesis, in the context of our model and collider bound 
on the mass of new gauge boson $Z^\prime$ is presented in Sec. \ref{sec:Zmass}. Finally, in  Sec. \ref{sec:conclude}, we conclude our results.

\section{Model Framework }
\label{sec:2}
In order to fulfil our desired goal, we incorporate new particles and assign them suitable charges under extended symmetries (i.e.,  modular  $A_4$ and $U(1)_{B-L}$), as presented in Table-\ref{tab:fields-type3}, such that the superpotential remains invariant.  The idea behind the inclusion of $U(1)_{B-L}$ symmetry along with $A_4$ modular symmetry is to avoid certain unwanted terms in the superpotential which is not possible by $A_4$ modular symmetry.  The suitability to go beyond standard model (BSM) paves the way to include heavy  RH neutrinos  $\Sigma_R$ in our model, which transform as triplet under  $SU(2)_L$,  and accompanying these, we have also included a weighton ($\rho$).  These symmetries are broken at a very high scale, much greater than the scale of electroweak symmetry breaking. The $U(1)_{B-L}$ symmetry is spontaneously broken by assigning non-zero VEV to the singlet weighton $\rho$ and the $Z^\prime$ boson associated with it acquires its mass by the singlet VEV $v_\rho$. We will show in sec.\ref{sec:Zmass} that its mass and gauge coupling satisfy the present experimental bounds. Moreover, the non-zero VEV acquired by the singlet weighton helps heavy RH neutrinos to gain mass. We implement modular symmetry because it restricts the usage of excess flavon fields, which otherwise, overfill the particle gamut and reduces the predictability of the model while working in BSM. This becomes possible only because Yukawa couplings acquire modular form and also takeover the job performed by extra flavon fields. In addition, the complete superpotential of our model is represented below,
\begin{table}[h!]
\begin{center}
\begin{tabular}{|c||c|c|c|c|c||c|c|} \hline\hline 

Fields & ~$E^c_{R_1}$~& ~$E^c_{R_2}$~  & ~$E^c_{R_3}$~& ~$L$~& ~$\Sigma^c_{Ri}$~&~ $H_{u,d}$~&~$\rho$ \\ \hline 
$SU(2)_L$ & $1$  & $1$  & $1$  & $2$  & $3$  & $2$ &$1$       \\\hline 
$U(1)_Y$   & $1$ & $1$ & $1$ & $-\frac{1}{2}$  & $0$ & $\frac12$,$-\frac12$  & $0$  \\\hline
 
$U(1)_{B-L}$   & $1$ & $1$ & $1$ & $-1$  & $1$ & $0$   & $-2$ \\\hline
$A_4$ & $1$ & $1'$ & $1''$ & $1, 1^{\prime \prime}, 1^{\prime }$ & $3$ & $1$ & $1$ \\ \hline
$k_I$ & $0$ & $0$ & $0$ & $0$ & $-2$ & $0$ & $2$  \\ \hline

\hline
\end{tabular}
\caption{Particle content of the model and their charges under ${ SU(2)_L\times U(1)_Y\times U(1)_{B-L}}\times A_4 $, where, $k_I$ is the modular weight.}
\label{tab:fields-type3}
\end{center}
\end{table}
\begin{table}[h!]
\begin{center}
\begin{tabular}{|c|c|c|}\hline\hline  
 Yukawa couplings~ & \multicolumn{1}{c||}{$A_4$} & \multicolumn{1}{c|}{$k_I$}\\ \hline \hline

$Y=(y_1,y_2,y_3)$ & ~$3$~& ~$2$~ \\ \hline 
\end{tabular}
\caption{ Charge assignment to Yukawa coupling under $A_4$ and its modular weight. }
\label{tab:Yukawa}
\end{center}
\end{table} \
\begin{eqnarray}
\mathcal{W}_{III} &= \mathcal{W}_{M_\ell} + \mathcal{W}_{M_D} + \mathcal{W}_{M_R},
\end{eqnarray}
where, the terms $\mathcal{W}_{M_\ell}$,  $\mathcal{W}_{M_D}$ and  $\mathcal{W}_{M_R}$ are responsible for generating the mass term  for  the charged leptons, Dirac mass term for the neutrinos and Majorana mass  term for the RH neutrinos  and  their explicit forms are provided in the following  subsections.
\vspace{0.5cm}
 \\
\underline{\textbf{Masses of charged leptons}}
\vspace{4mm}

We urge to have a simplified form of charged lepton mass matrix for which we assign $U(1)_{B-L}$ charge to the right-handed (RH) charged leptons i.e., $E_{Ri}^c$ as $+1$, and three generations of left-handed (LH) charged leptons have the value   $-1$. While under $A_4$ symmetry RH and LH charged leptons transform as $\{1, 1^\prime, 1^{\prime \prime}\}$ and \{$1, 1^{\prime \prime}, 1^\prime$\}. In addition, the modular weight assigned to the charged leptons is zero. The Higgsinos $H_{u,d}$ are given charges $0$ and $1$ under $U(1)_{B-L}$ and $A_4$ symmetry respectively, with zero modular weight. The VEVs of Higgsinos i.e., $(v_u, v_d)$ are related to the SM Higgs VEV $(v_H)$ by a simple equation $v_H =\frac12 \sqrt{v_u^2 + v_d^2}$. The ratio of Higgsinos VEV is written as $\tan\beta = ({v_u}/{v_d}) \simeq 5$ (used in our analysis) \cite{Antusch:2013jca,Okada:2019uoy,Bjorkeroth:2015ora}. The admissible superpotential term for the charged lepton sector is given below:
\begin{eqnarray}
 \mathcal{W}_{M_\ell}  
                   &=  y_{ij}E^c_{R_i}H_d L_j \;. 
                    \label{charged lepton}
\end{eqnarray}
After the electroweak symmetry breaking the mass matrix for the charged leptons takes the diagonal form:
\begin{align}
M_\ell&=\frac{v_d}{\sqrt2}
\left[\begin{array}{ccc}
y_{ee} & 0 & 0 \\ 
0 & y_{\mu\mu} & 0 \\ 
0 & 0 & y_{\tau\tau} \\ 
\end{array}\right]
.     
\end{align}
\vspace{1cm}\\
\underline{\textbf{Dirac mass term}}\\

The neutral lepton sector gets mass as and when $H_u$ acquires non-vanishing VEV. To keep Dirac term invariant under $A_4$ modular group, we need fermion triplets to have charge $3$ as Yukawa couplings are triplet (\textbf{Y}=$(y_1,y_2,y_3$)). Hence, the Dirac interaction term of neutral multiplet of fermion triplet with the SM left-handed neutral leptons can be written as:
\begin{eqnarray}
 \mathcal{W}_{M_D}  
                   &=   - ~ (G_D)_{ij} \left[{H_u} \Sigma^c_{R_i} \sqrt{2}\textbf{Y} L_j \right]  \;,  \label{eqn:Dirac1}
\end{eqnarray}
with $G_D $= diag\{$\alpha_1,\alpha_2, \alpha_3$\}, which gives the mass matrix
\begin{align}
M_D&=v_u
\left[\begin{array}{ccc}
\alpha_1 & 0 & 0 \\ 
0 & \alpha_2 & 0 \\ 
0 & 0 & \alpha_3 \\ 
\end{array}\right]
\left[\begin{array}{ccc}
y_1 &y_3 &y_2 \\ 
y_2 &y_1 &y_3 \\ 
y_3 &y_2 &y_1 \\ 
\end{array}\right].     \label{eqn:Dirac}
\end{align}\
\underline{\textbf{Majorana mass term}}\
\vspace{4mm}

The superpotential for Majorana mass term for right handed neutrinos is given as,
\begin{equation}
\mathcal{W}_{M_R}= -\frac{M^{\prime}_{\Sigma_{}}}{2} \left(\beta_{\Sigma} {\rm{Tr}}  \left[\mathbf{{\Sigma^c_{R_i}}}\textbf{Y} \mathbf{{\Sigma^c_{R_i}}}\right]_{\rm{sym}} + \gamma_\Sigma {\rm{Tr}} \left[\mathbf{{\Sigma^c_{R_i}}}\textbf {Y} \mathbf{ {\Sigma^c_{R_i}}}\right]_{\rm asym}\right) \frac{\rho}{\Lambda} \;,
\label{Maj_matrix}
\end{equation}
where, $M^{\prime}_{\Sigma_{}}$ is the free mass parameter and {$\Sigma^c_{R_i}$ with $(i=1,2,3)$}, which can be represented in $SU(2)$ basis as,
\begin{eqnarray}
\Sigma^c_{R_i} = \begin{pmatrix}
\Sigma^{0~c}_{R_i}/\sqrt{2} ~&~ \Sigma^{-~c}_{R_i}\\
\Sigma^{+~c}_{R_i} ~&~ -\Sigma^{0~c}_{R_i}/\sqrt{2} 
\end{pmatrix}.
\end{eqnarray}
 Applying $A_4$ symmetry product rule to eqn. (\ref{Maj_matrix}), yields  both symmetric and anti-symmetric parts with $\beta_{\Sigma}$ = diag\{$\beta_1, \beta_2, \beta_3$\} and $\gamma_{\Sigma}$ = diag\{$\gamma_1, \gamma_2, \gamma_3$\} being the associated free parameter matrices respectively:
\begin{align}
M_{R}&=\frac{v_\rho}{\Lambda \sqrt2} \left(\frac{M^{\prime}_{\Sigma}}{2}\right) \left( \frac{\beta_{\Sigma}}{3}\left[\begin{array}{ccc}
2y_1 & -y_3 & -y_2 \\ 
-y_3 & 2y_2 & -y_1 \\ 
-y_2 & -y_1 & 2y_3 \\ 
\end{array}\right] + \gamma_{\Sigma} \left[\begin{array}{ccc}
0 & y_3 & -y_2 \\ 
-y_3 & 0 & y_1 \\ 
y_2 & -y_1 & 0 \\ 
\end{array}\right]\right).
\label{MR_matrix}
\end{align}\
The active neutrino mass matrix in the framework of type-III seesaw is given as,
\begin{align}
m_\nu&= - M_D M_R^{-1} M_D^T\;.
\label{nmass}
\end{align}
\vspace{0.25cm}
\section{Numerical Analysis}
\label{sec:numerical}
\begin{figure}[htpb]
\begin{center}
\includegraphics[height=50mm,width=75mm]{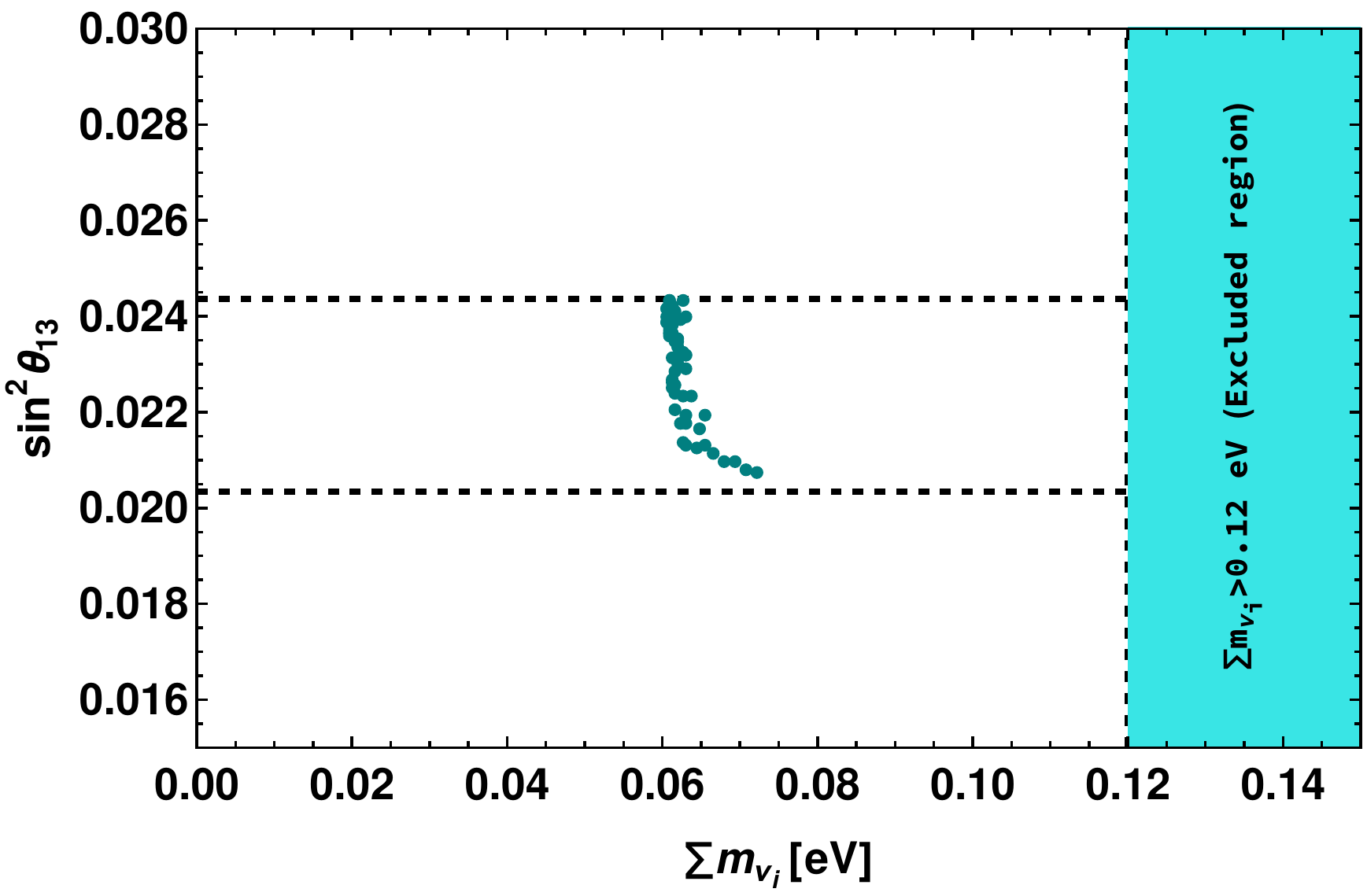}
\hspace*{0.2 true cm}
\includegraphics[height=50mm,width=75mm]{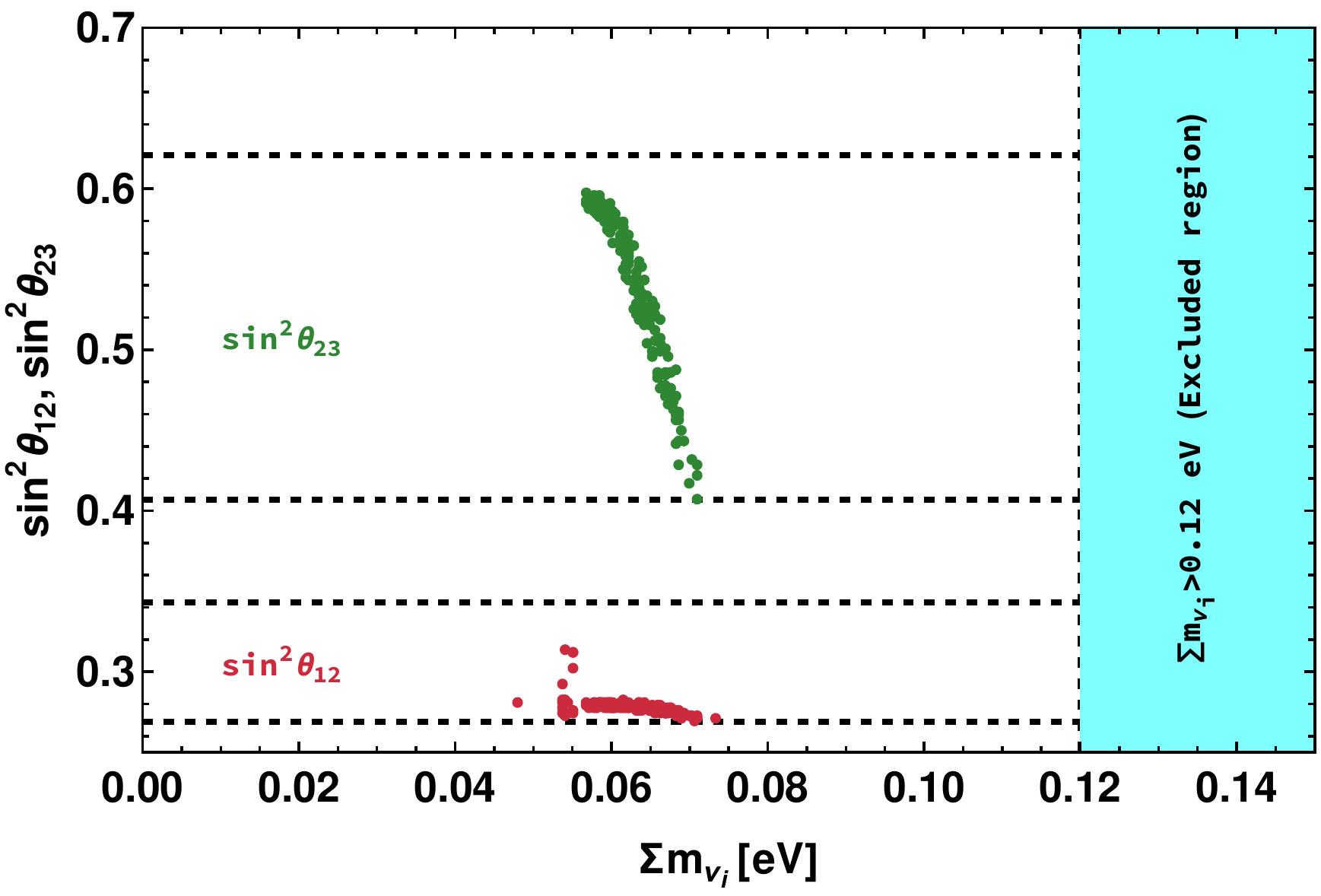}
\caption{Left(right) panel shows the  plane of  the mixing angle i.e., $\sin^2 \theta_{13}$~($\sin^2 \theta_{12},\sin^2 \theta_{23}$) with sum of neutrino mass for the best fit values of model parameters while grid-lines represent the $3\sigma$ range of mixing angles.}
\label{correl_mix_sum}
\end{center}
\end{figure}

\begin{figure}[htpb]
\begin{center}
\includegraphics[height=50mm,width=75mm]{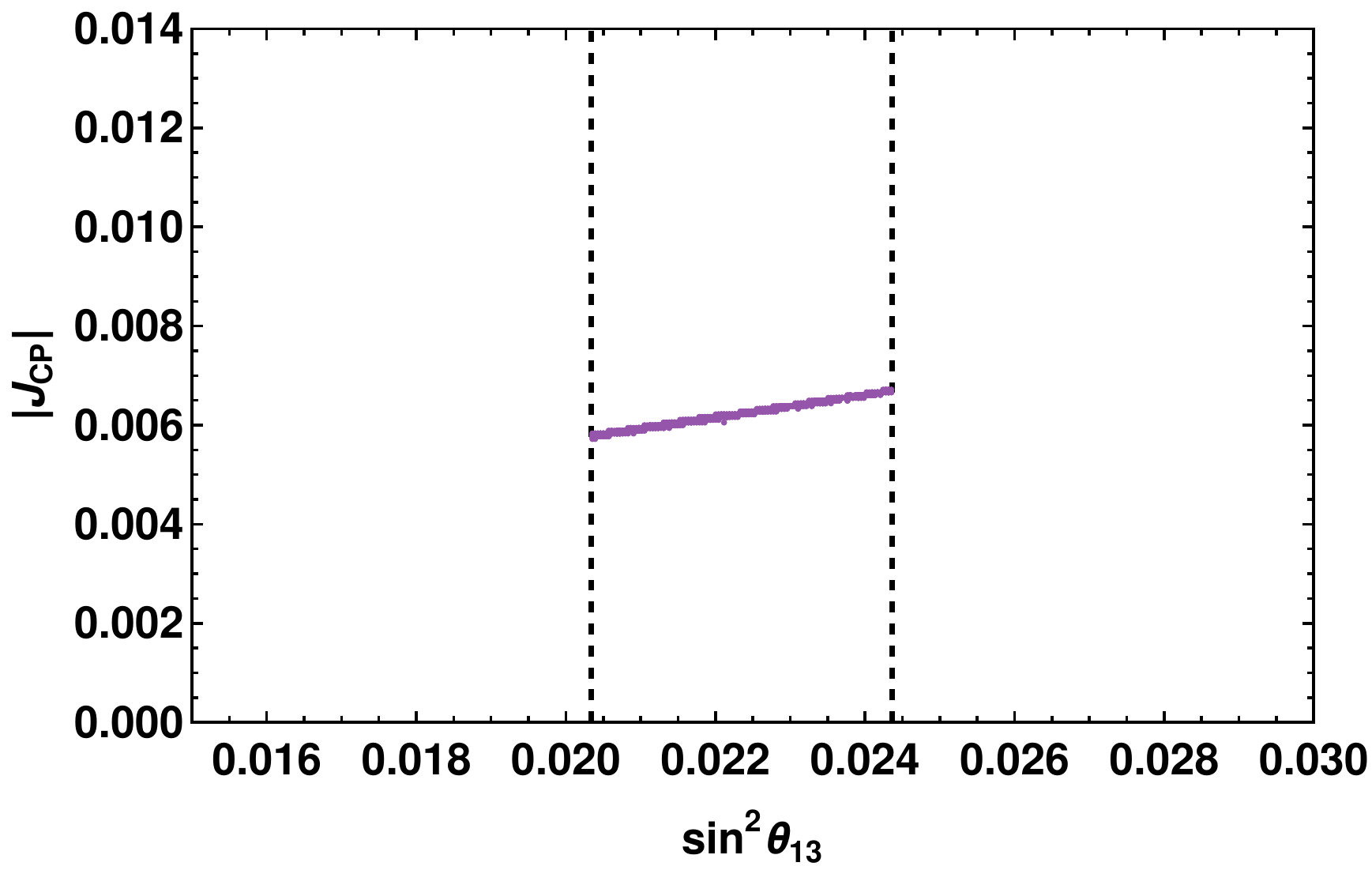}
\hspace*{0.2 true cm}
\includegraphics[height=50mm,width=75mm]{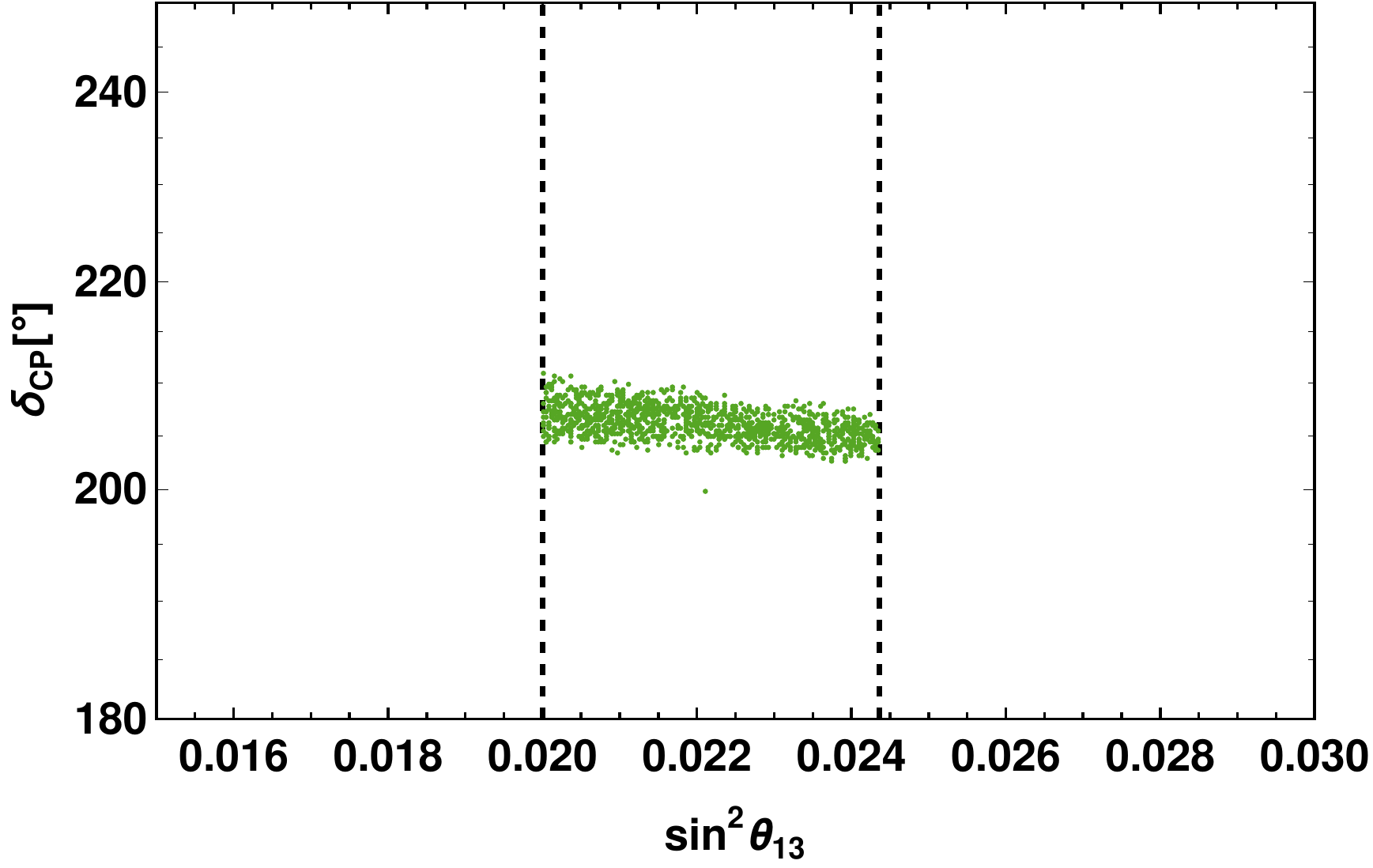}
\caption{Left (right) panel shows the correlation of $\sin^2 \theta_{13}$ with $J_{CP}$ ($\delta_{CP}$).}
\label{jcp_dcp}
\end{center}
\end{figure}

The global fit neutrino oscillation data at 3$\sigma$ interval from \cite{Esteban:2020cvm} is used for numerical analysis, as given in Table \ref{table:expt_value}.
\begin{table}[htbp]
\centering
\begin{tabular}{|c|c|c|c|c|c|c|}
\hline
\bf{Oscillation Parameters} & \bf{Best fit value} \bf{$\pm$ $1\sigma$} & \bf{ 2$\sigma$ range}& \bf{3$\sigma$ range} \\
\hline \hline
$\Delta m^2_{21}[10^{-5}~{\rm eV}^2]$& 7.56$\pm$0.19  & 7.20--7.95 & 7.05--8.14  \\
\hline
$|\Delta m^2_{31}|[10^{-3}~{\rm eV}^2]$ (NO) &  2.55$\pm$0.04 &  2.47--2.63 &  2.43--2.67\\
\hline
$\sin^2\theta_{12} / 10^{-1}$ & 3.21$^{+0.18}_{-0.16}$ & 2.89--3.59 & 2.73--3.79\\
\hline
$\sin^2\theta_{23} / 10^{-1}$ (NO)
	  &	4.30$^{+0.20}_{-0.18}$ 
	& 3.98--4.78 \& 5.60--6.17 & 3.84--6.35 \\
	  & 5.98$^{+0.17}_{-0.15}$ 
	& 4.09--4.42 \& 5.61--6.27 & 3.89--4.88 \& 5.22--6.41 \\
\hline
$\sin^2\theta_{13} / 10^{-2}$ (NO) & 2.155$^{+0.090}_{-0.075}$ &  $1.98-2.31$ & $2.04-2.43$ \\
\hline 
$\delta_{CP} / \pi$ (NO) & 1.08$^{+0.13}_{-0.12}$ & $0.84 - 1.42$ & $0.71 - 1.99$\\
\hline
\end{tabular}
\caption{The global-fit values of the  oscillation parameters along with their 1$\sigma$/2$\sigma$/3$\sigma$ ranges.}
\label{table:expt_value}
\end{table}
The neutrino mass matrix calculated using eqn.(\ref{nmass}) is numerically diagonalized using the relation $U^\dagger \mathcal{M} U= {\rm diag}(m_{\nu_1}^2, m_{\nu_2}^2, m_{\nu_3}^2)$, where, ${\cal M}=m_\nu m_\nu^\dagger$ and $U$ is a unitary matrix, from which the neutrino mixing angles can be derived using the conventional relations:
\begin{eqnarray}
\sin^2 \theta_{13}= |U_{13}|^2,~~~~\sin^2 \theta_{12}= \frac{|U_{12}|^2}{1-|U_{13}|^2}\;,~~~~~\sin^2 \theta_{23}= \frac{|U_{23}|^2}{1-|U_{13}|^2}\;.
\end{eqnarray}
Other observables related  to the mixing angles and phases of PMNS matrix are
\begin{eqnarray}
J_{CP} &=& \text{Im} [U_{e1} U_{\mu 2} U_{e 2}^* U_{\mu 1}^*] = s_{23} c_{23} s_{12} c_{12} s_{13} c^2_{13} \sin \delta_{CP}\;,\\
\langle m_{ee}\rangle &=&|m_{\nu_1} \cos^2\theta_{12} \cos^2\theta_{13}+ m_{\nu_2} \sin^2\theta_{12} \cos^2\theta_{13}e^{i\alpha_{21}}+  m_{\nu_3} \sin^2\theta_{13}e^{i(\alpha_{31}-2\delta_{CP})}|.
\end{eqnarray}
The effective Majorana mass parameter $\langle m_{ee}\rangle$ is expected to have improved sensitivity measured by KamLAND-Zen experiment in coming future~\cite{KamLAND-Zen:2016pfg}. 
Further, we chose the following model parameter ranges to fit the present neutrino oscillation data:
\begin{eqnarray}
{\rm Re}[\tau] \in [-0.5,0.5],~~{\rm Im}[\tau]\in [0.75,2],
\quad M^{\prime}_\Sigma \in  [10^4,10^5] \ {\rm TeV} ,\quad v_\rho \in  [10^3,10^4] ~{\rm TeV}, \nn \\ \Lambda \in  [10^4,10^5]  \ {\rm TeV},~~ G_D \in  [10^{-8},{10^{-5}}],~~ \beta_{\Sigma} \in  [{10^{-5}},10^{-1}],~~ \gamma_{\Sigma} \in  [10^{-9},10^{-10}].
\label{ranges}
\end{eqnarray}
We consider the free mass parameter ($M^\prime_\Sigma$), real and imaginary part of $\tau$, VEV of weighton ($v_\rho$) and cut-off parameter ($\Lambda$) randomly in the range given in eqn.(\ref{ranges}). 
The range of $\tau$ is taken to be $[-0.5,0.5]$ for the real part and $[0.75,2]$ for the imaginary part, which provides the validity of model to follow normal hierarchy (NH). Considering these ranges, we arbitrarily scrutinise the input values of parameters and extract the best-fit values of those by applying chi-square minimization technique. The approach followed here by considering the general chi-square formula \cite{Roe:2015fca,Ding:2021eva}, which is utilized for calculating the $\chi^2$ values for all the available observables of the neutrino sector, like two mass squared differences and three mixing angles, further yielding cumulative $\chi^2$ minimum allowing us to get the values of the free parameters corresponding to the minimum i.e., best-fit values \cite{Wang:2020lxk,Novichkov:2020eep}. As there are a large number of free parameters involved in this framework, i.e., total number  of free parameters are  much larger than the number of observed neutrino oscillation parameters,  it is not possible to get a  constrained a correlated plot. Therefore, we calculate minimal chi-square and the the associated values  of free parameters are considered as the best-fit values of  the free parameters. Hence, Table \ref{tab:bestfit} is obtained by keeping the experimentally observed oscillation parameters along with the cosmological bound  for sum of neutrino masses $\Sigma m_{\nu_i} \leq 0.12 ~\rm eV$ \citep{Planck:2018vyg}.  
\begin{table}
\centering
\begin{tabular}{|c|c|c|c|c|c|c|}\hline\hline  
 Model Parameters & \multicolumn{1}{c|}{~$\alpha_1$~} & \multicolumn{1}{c|}{$\alpha_2$} & \multicolumn{1}{c|}{$\alpha_3$}& \multicolumn{1}{c|}{$\beta_1$} & \multicolumn{1}{c|}{$\beta_2$} & \multicolumn{1}{c|}{$\beta_3$}\\ \hline

Best-fit values& ~$3.83 \times 10^{-7}$~& ~$1.61 \times 10^{-6}$~&$5.73 \times 10^{-7}$&$4.44 \times 10^{-2}$&$~~~~0.824~~~~$ & $1.05 \times 10^{-3}$ ~\\ \hline 
\end{tabular}
\caption{Best-fit of model parameters by under constraint of experimentally observed data.}
\label{tab:bestfit}
\end{table} 
We have not mentioned the best-fit values of $\gamma_\Sigma$ here in the Table \ref{tab:bestfit}, since it gives negligible contribution to the observables as compared to $G_D$ and $\beta_\Sigma$, hence, conventionally we deal with total six free parameters. As a consequence, the left panel of Fig.\ref{correl_mix_sum} projects the correlation between $\sin^2\theta_{13}$ w.r.t. $\sum m_{\nu_i}$, where, the sum of neutrino mass is above its lower bound i.e., $0.058$eV \citep{RoyChoudhury:2019hls}, while the right panel shows the inter-dependence of $\sum m_{\nu_i}$ with  ($\sin^2 \theta_{12}, \sin^2 \theta_{23}$) with grid-lines showing their respective $3\sigma$ ranges. Moreover, in Fig. \ref{jcp_dcp} the left panel shows an interdependence of $\sin^2\theta_{13}$ with Jarlskog invariant $| J_{CP} |$  whose value is constrained to be $| J_{CP}| \leq 7.1 \times 10^{-3}$. 
As can be seen in the plot of $\delta_{CP}$ against $\sin^2\theta_{13}$ on the right panel of Fig. \ref{jcp_dcp}, $\delta_{CP}$ is varying in the range $[202^\circ$ - $211^\circ]$ while constrained by 3$\sigma$ bound of  $\sin^2\theta_{13}$. 

\begin{figure}[htpb]
\begin{center}
\includegraphics[height=55mm,width=58mm]{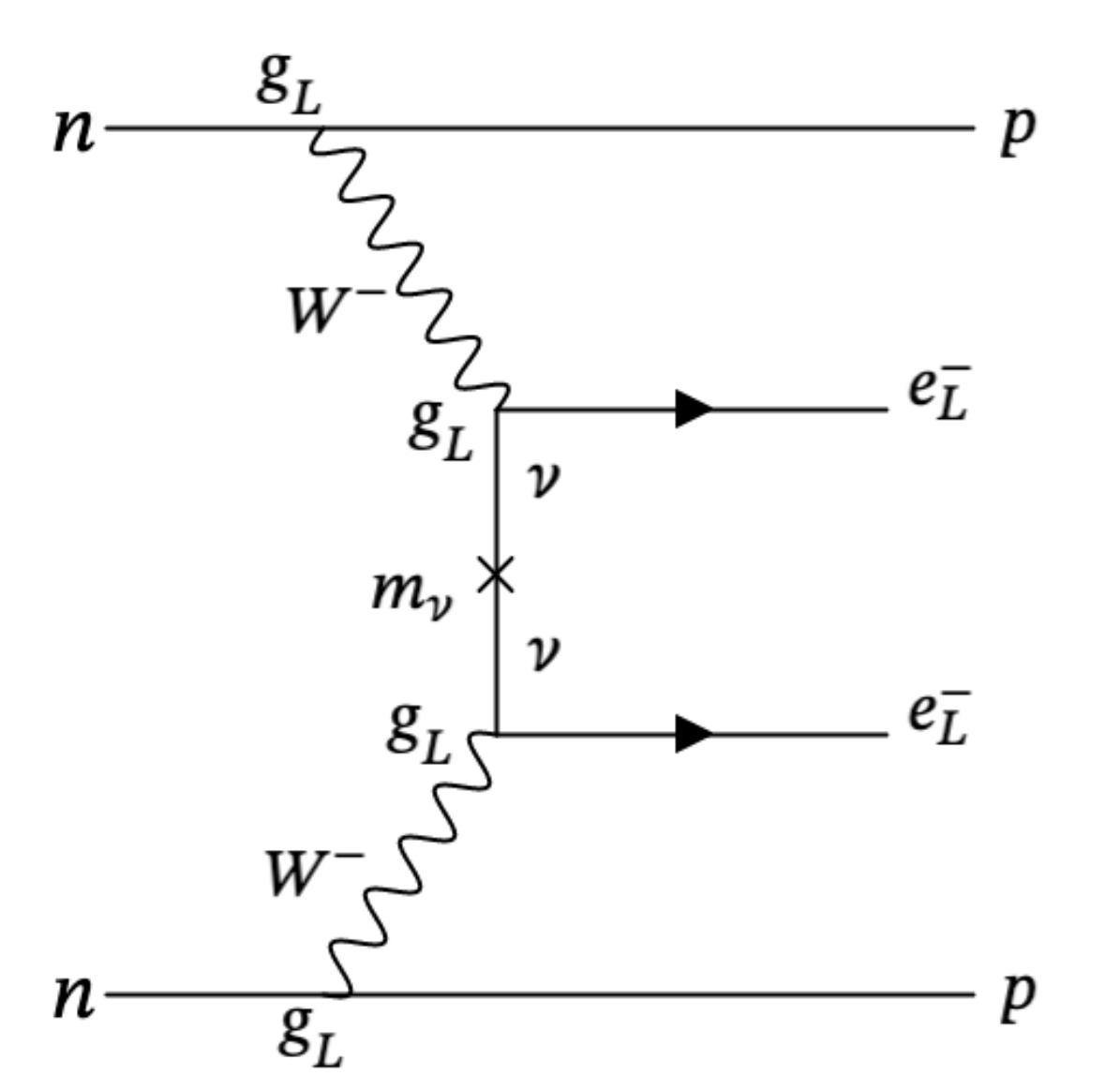}
\hspace{9mm}
\includegraphics[height=55mm,width=58mm]{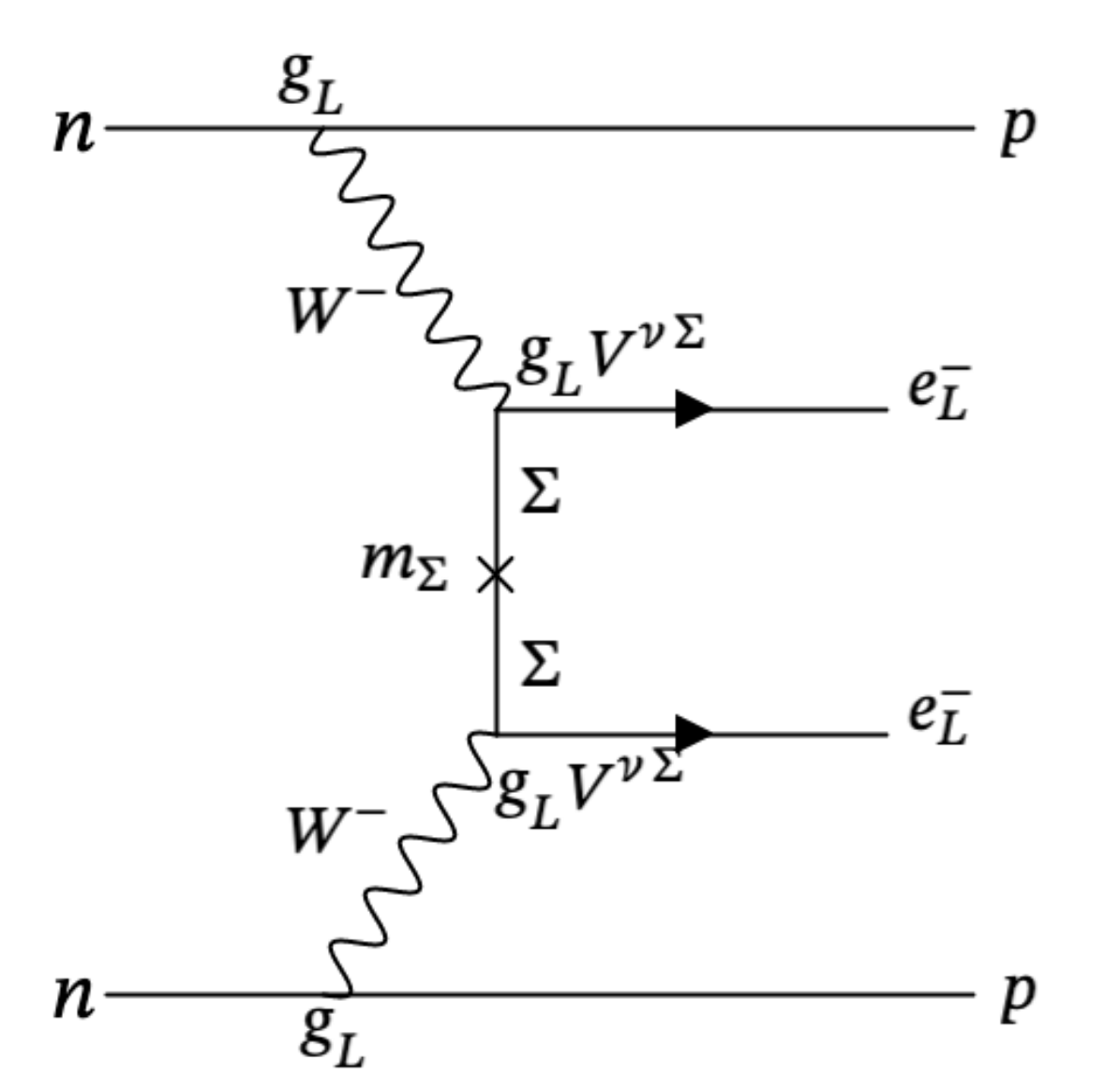}
\caption{Feynman diagrams contributing to neutrinoless double beta decay with
$W^--W^-$ mediation via the exchange of virtual light neutrinos $\nu$ (left panel), and the exchange of virtual heavy neutrinos $\Sigma$ (right panel), where $g_L$ represents the SM weak coupling.}
\label{NDBD_SM_BSM}
\end{center}
\end{figure}
The process of neutrinoless double beta decay (NDBD) involves the simultaneous conversion of two neutrons into two protons and two electrons without any emission of neutrinos  \cite{Nucciotti:2007jk,Robertson:2013ziv,Cardani:2018lje,Dolinski:2019nrj}, as shown in the left panel of Fig. \ref{NDBD_SM_BSM}. In the presence of new heavy neutral fermions, the additional  contribution to the NDBD is linked to the mixing between active and heavy neutrinos and is expected to be rather small. 
The  mixing of active and sterile neutrinos is generally descibed by the parameters $U_{\alpha i}$ and $\Theta_{\alpha i}$ which plays a crucial role \cite{Asaka:2011pb} in the description of neutrinoless double beta decay. Thus, the $6 \times 6$ neutrino mass matrix takes the form
\begin{equation}
\hat{M}= \begin{pmatrix}
0 & M_D\\
M_D^T & M_\Sigma
\end{pmatrix}.
\end{equation}
We can diagonalize it by using the unitary matrix $\hat{U}$ as $\hat{U} ^\dagger \hat{M}  \hat{U}^* = \hat{M}^{diag} $. The seesaw mechanism shows that $\hat{U}$ at the leading order takes the form \cite{Asaka:2011pb}
\begin{equation}
\hat U =\begin{pmatrix}
U & \Theta\\
-\Theta^\dagger U & 1
\end{pmatrix},
\end{equation}
where, $U$ is the PMNS matrix, diagonalizing the light active neutrino mass matrix as
\begin{equation}
U^\dagger M_\nu U^* = {\rm diag}(m_1,m_2,m_3),
\end{equation}
with $M_\nu = - M_D M_\Sigma^{-1} M_D^T$ as the light neutrino mass matrix obtained from Type-III seesaw. The eigenstates related to masses $m_i$ and $M_\Sigma$ are $\nu_i$ and $\Sigma_{R_i}$. The neutrino mixing in the charged current is then induced through
\begin{equation}
\nu_{L\alpha} = U_{\alpha i} \nu_i + \Theta_{\alpha i} \Sigma_{R_i}
\end{equation}
where, the $3\times3$ mixing matrix $\Theta$ is found to at the leading order as
\begin{equation}
\Theta_{\alpha i} = \frac{\left[{M_D}\right]_{\alpha i}}{M_\Sigma}
\end{equation}
The  vertex coupling, thus given as  $V^{\nu \Sigma} = \frac{1}{v_u} M_D U^{-1} M_\nu^{-1}$ \cite{Dash:2021pbx}.

\begin{figure}[htpb]
\begin{center}
\includegraphics[height=50mm,width=75mm]{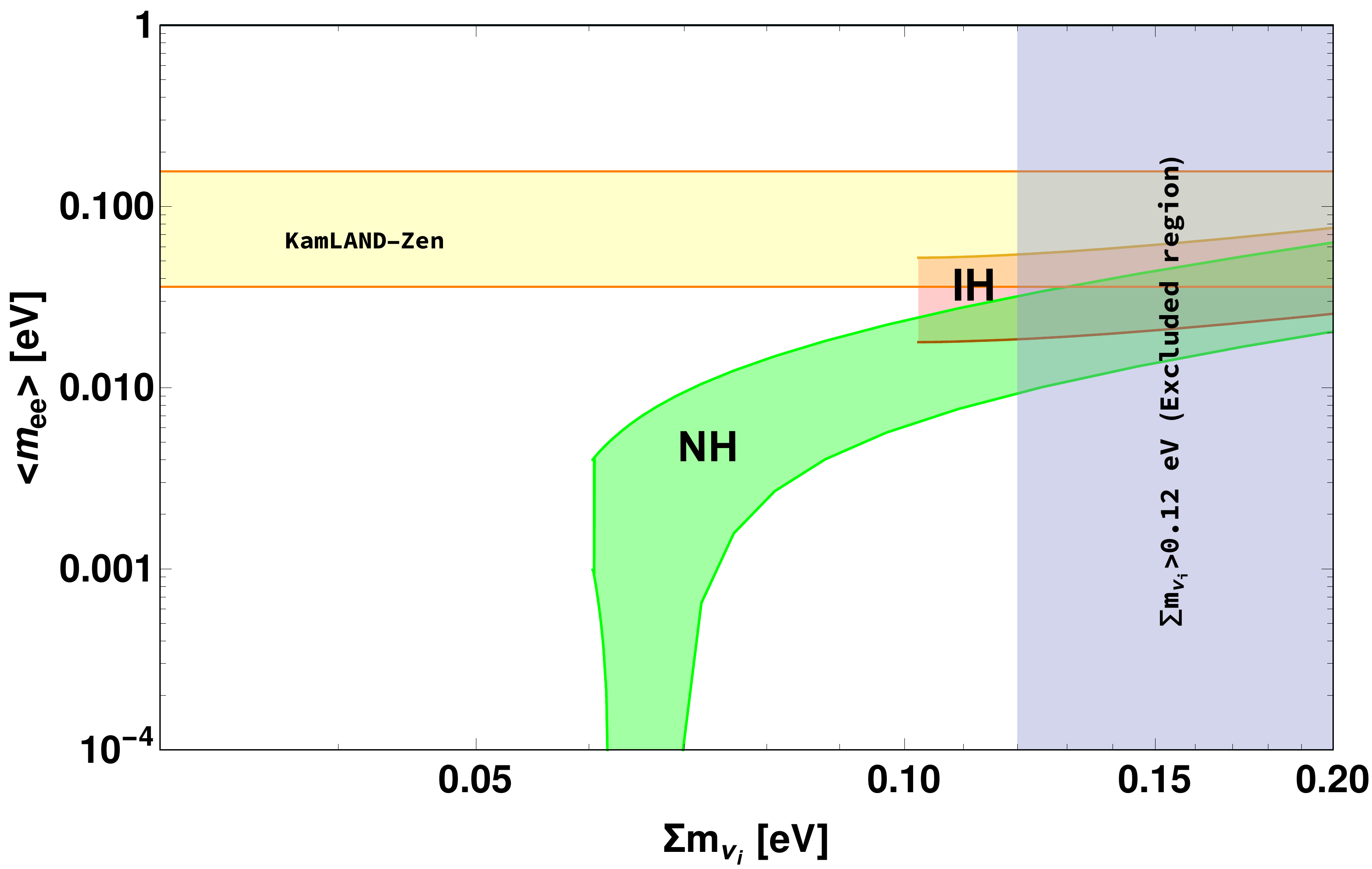}
\includegraphics[height=50mm,width=75mm]{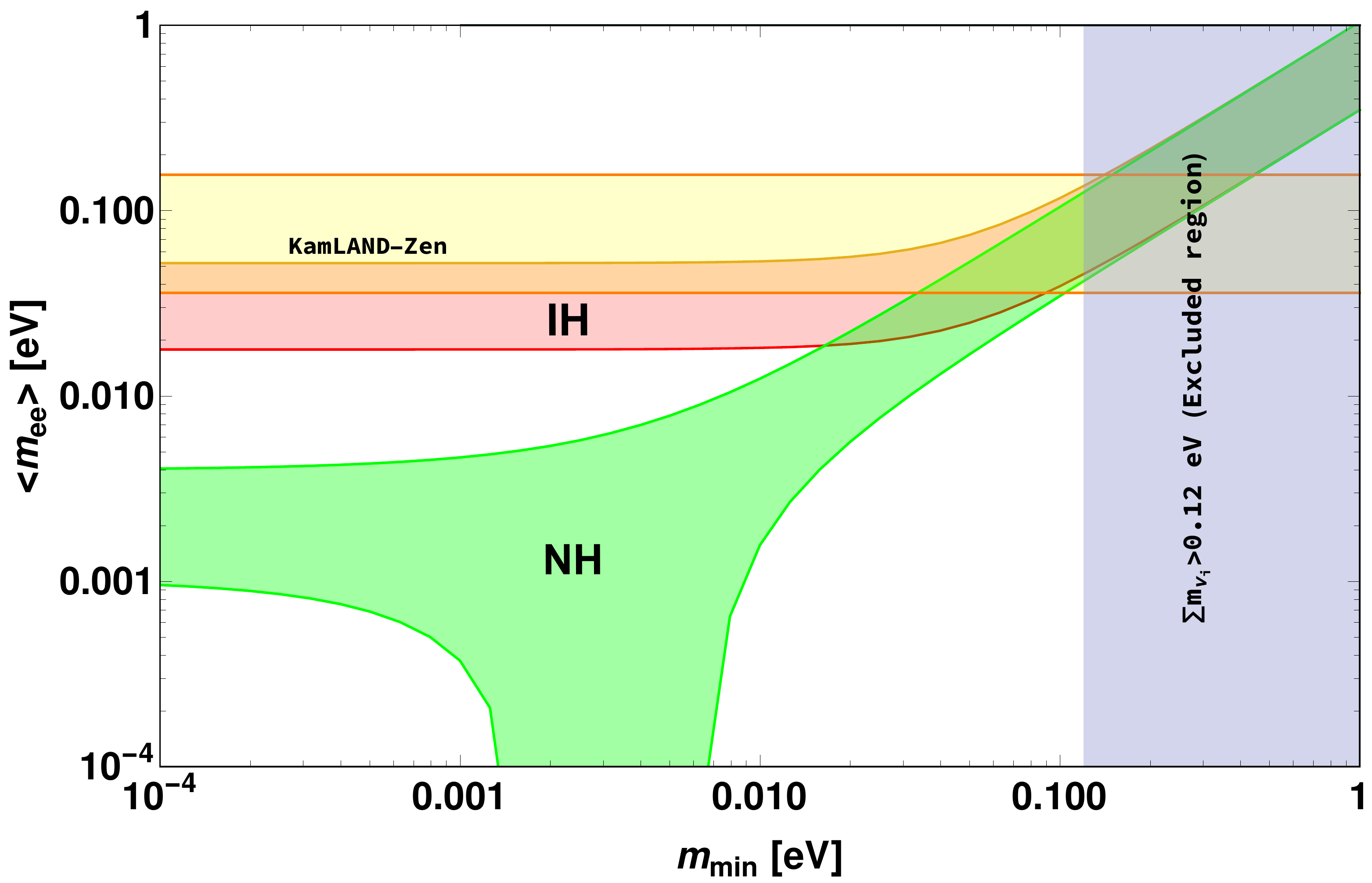}
\vspace{1cm}\\
\includegraphics[height=50mm,width=75mm]{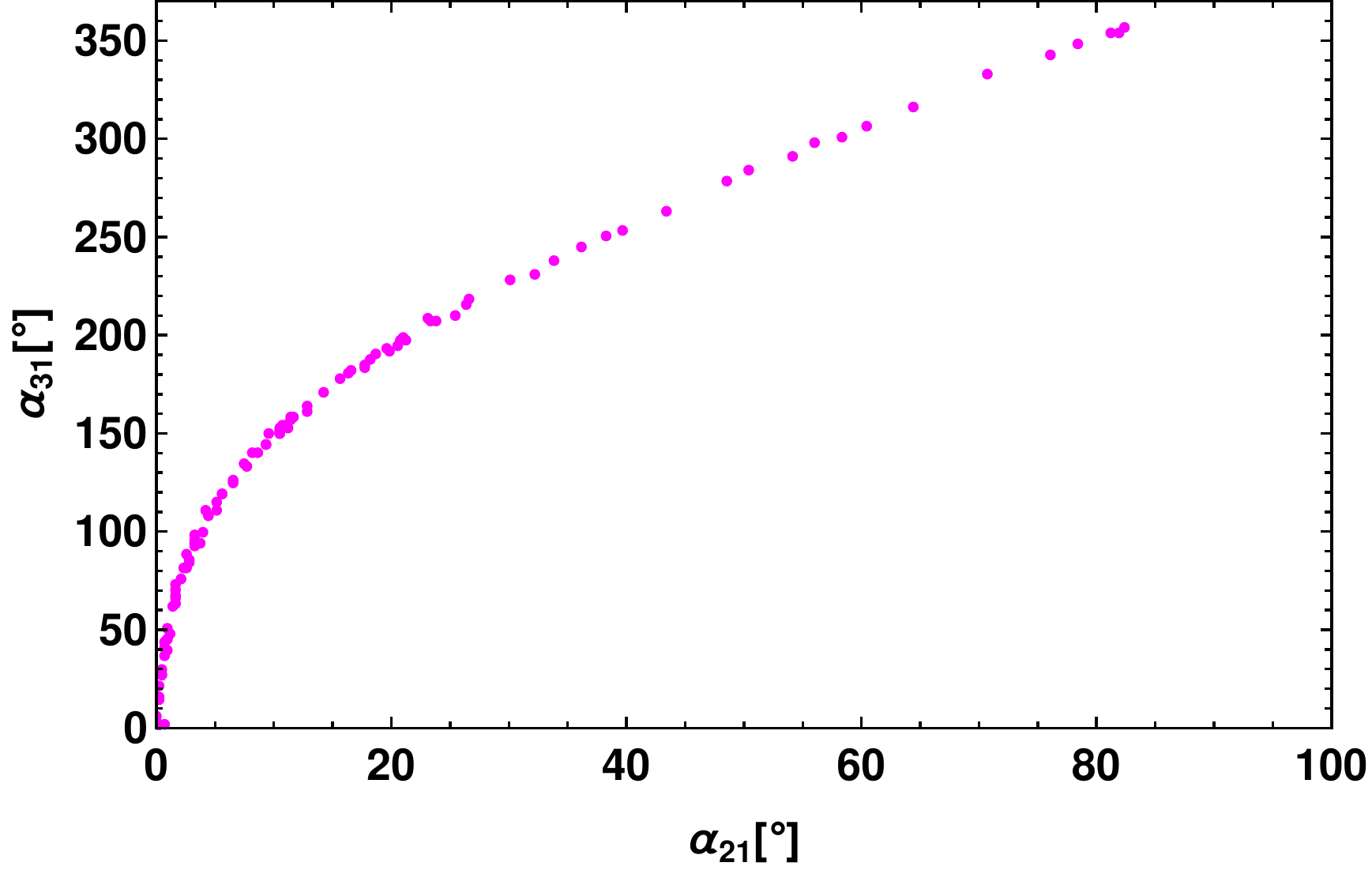}
\caption{Upper left panel shows the correlation in the  plane of  effective mass parameter $\langle m_{ee} \rangle$ and $\sum m_{\nu_i}$, whereas, upper right panel projects the correlation of $\langle m_{ee} \rangle$ and with lightest neutrino mass $m_1$ (i.e., $m_{min}$). Lower panel shows the correlation Majorana phases $\alpha_{21}$ and $\alpha_{31}$.}
\label{correl_mee_mix}
\end{center}
\end{figure}

 The right panel  of  Fig. \ref{NDBD_SM_BSM} represents the Feynman diagram due the exchange of the heavy neutrinos $\Sigma_{Ri}$ consisting all the relevant vertex couplings \cite{Dash:2021pbx} whose relevance is seen in numerical deductions, due to which the effective mass parameter $\langle m_{ee} \rangle$ receives additional contribution. We showcase the results in Fig. \ref{correl_mee_mix}, wherein  the upper left (right) panel reflects the behaviour  of $\langle m_{ee} \rangle$ \cite{Agostini:2022zub,King:2013psa} w.r.t. sum of neutrino mass ($\sum m_{\nu_i}$) (lightest neutrino mass $m_{\rm min})$ \cite{Gehrlein:2020jnr, Barry:2010yk} abiding the KamLAND-Zen bound \citep{KamLAND-Zen:2022tow} and the bottom panel shows the correlation between Majorana phases i.e., ($\alpha_{21}$ and $\alpha_{31}$).  \\

\begin{figure}[htpb]
\begin{center}
\includegraphics[height=50mm,width=75mm]{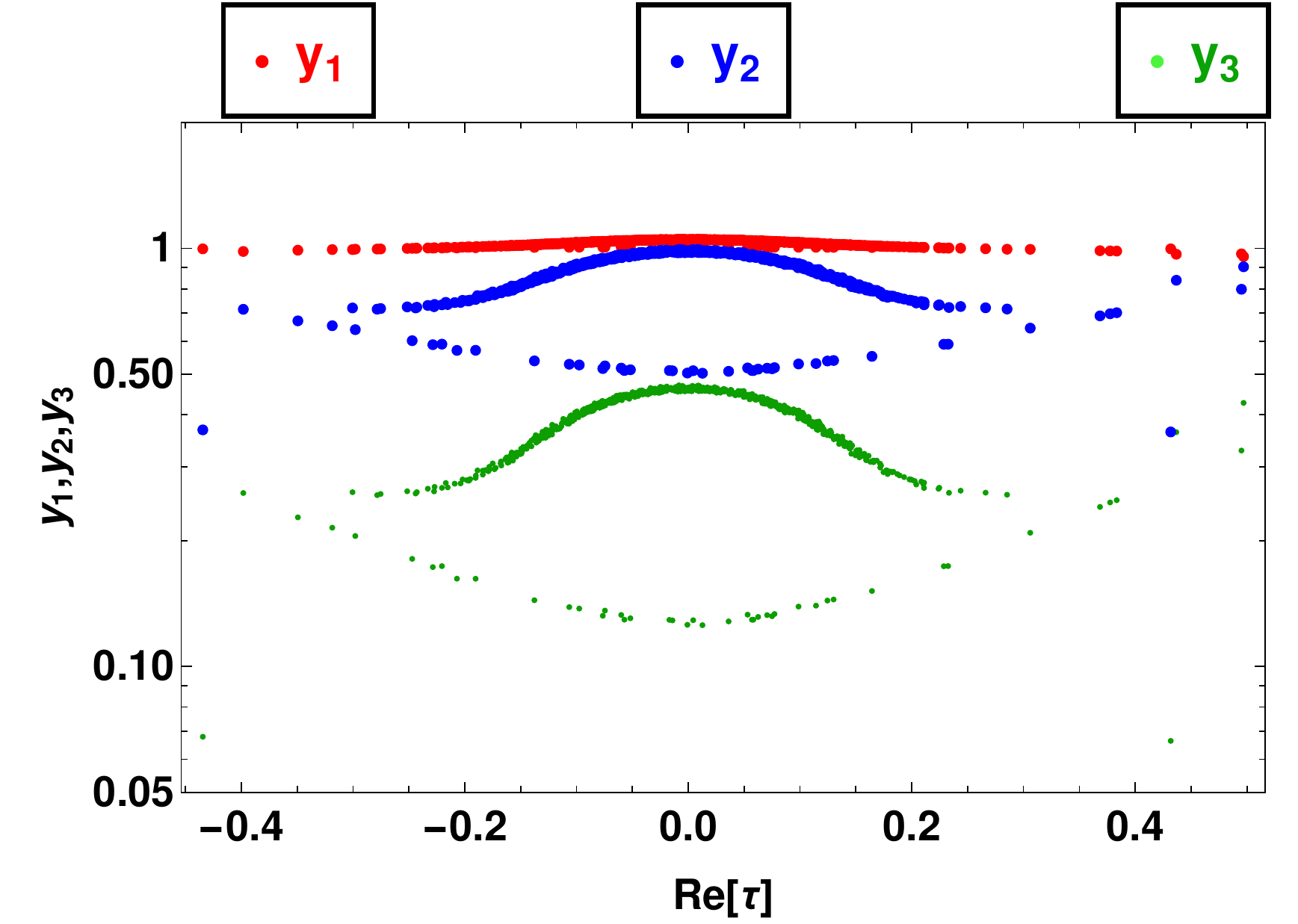}
\hspace*{0.2 true cm}
\includegraphics[height=50mm,width=75mm]{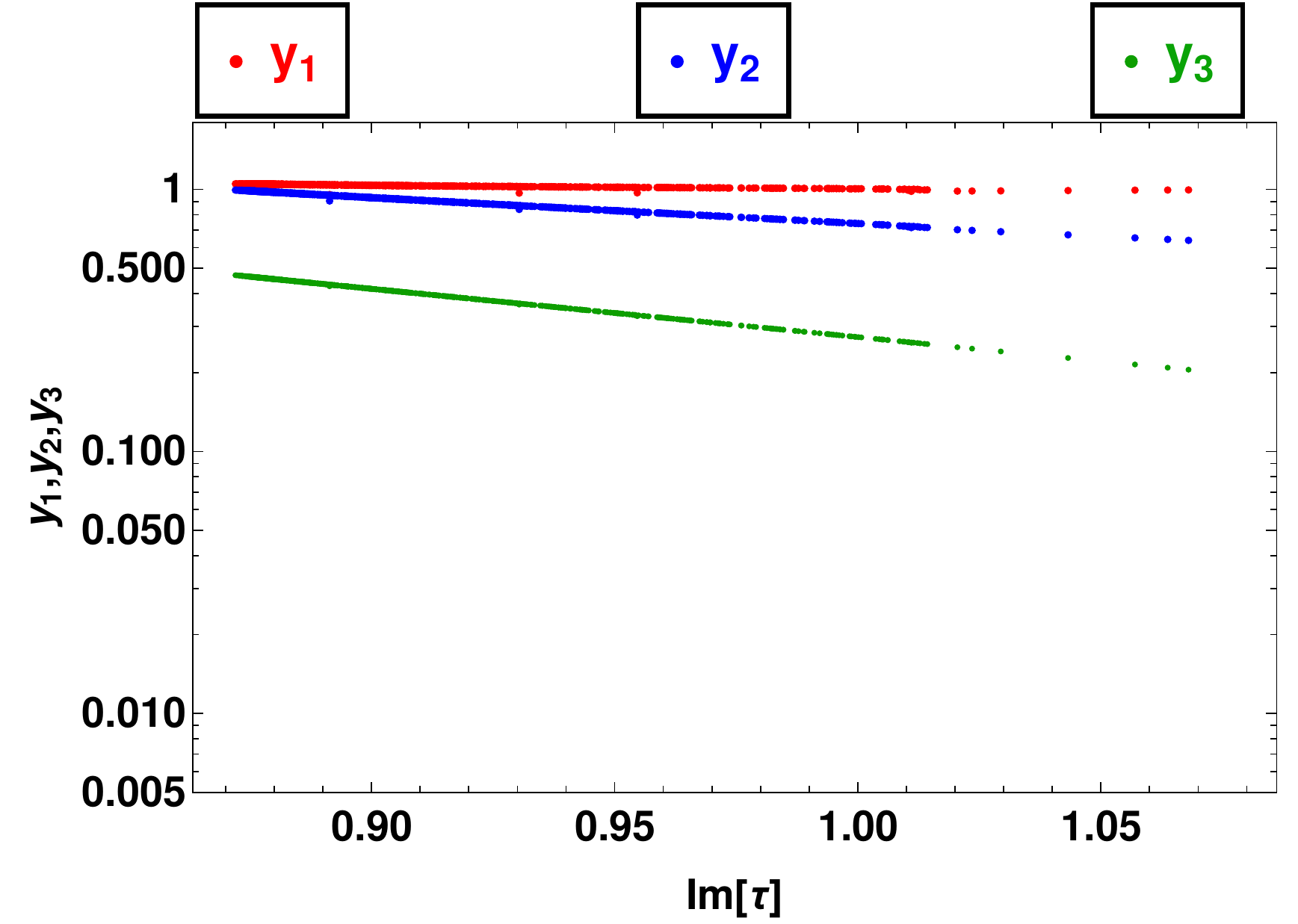}
\caption{Left (right) panel shows the correlation of the Yukawa couplings i.e., ($y_1, y_2, y_3$) w.r.t. Re[$\tau$] (Im[$\tau$]), where $y_1,y_2$ and $y_3,$ are shown in red, blue and green colours respectively.}
\label{yukawa}
\end{center}
\end{figure}

\begin{figure}[htpb]
\begin{center}
\includegraphics[height=50mm,width=75mm]{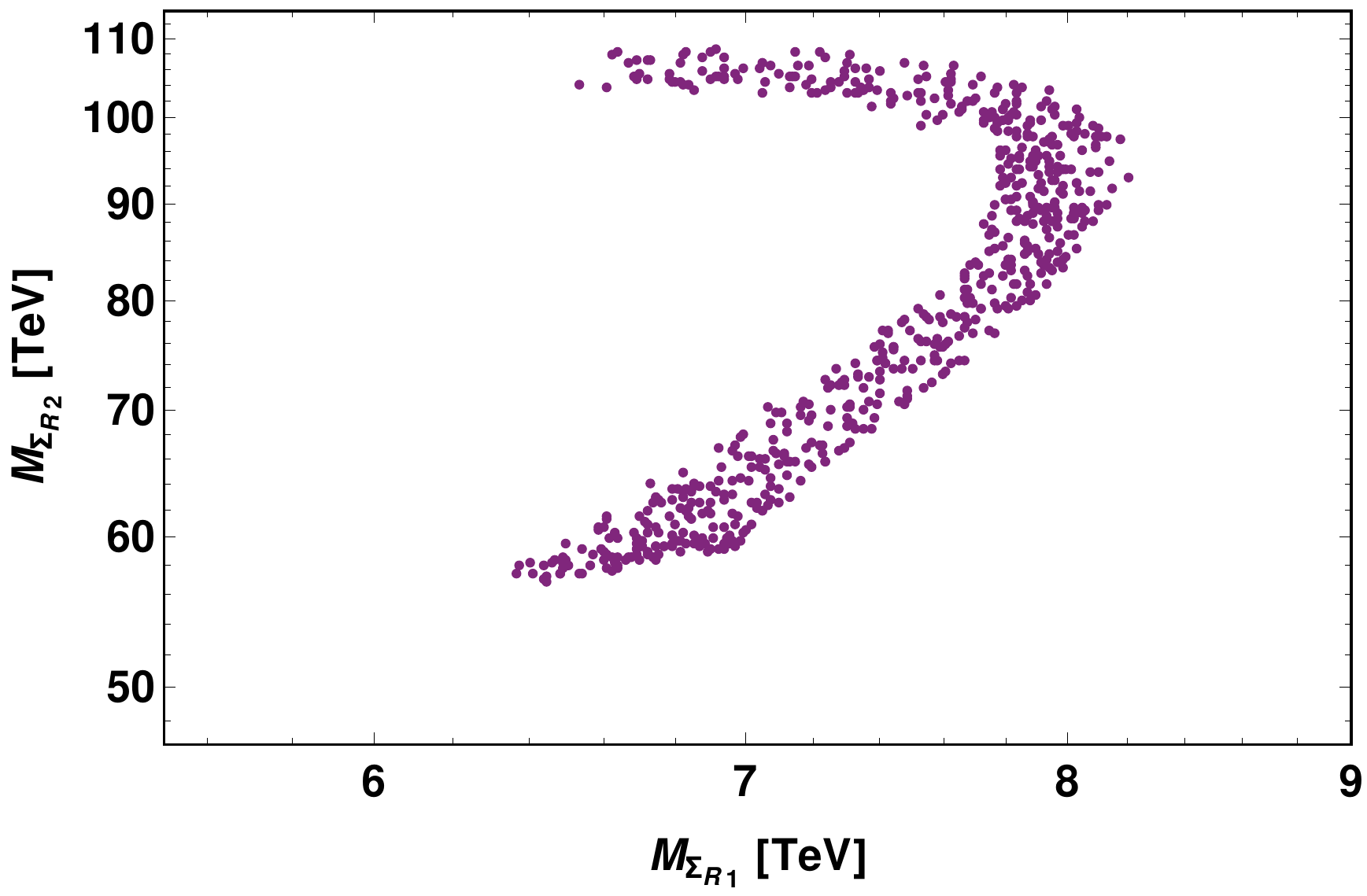}
\hspace*{0.2 true cm}
\includegraphics[height=50mm,width=75mm]{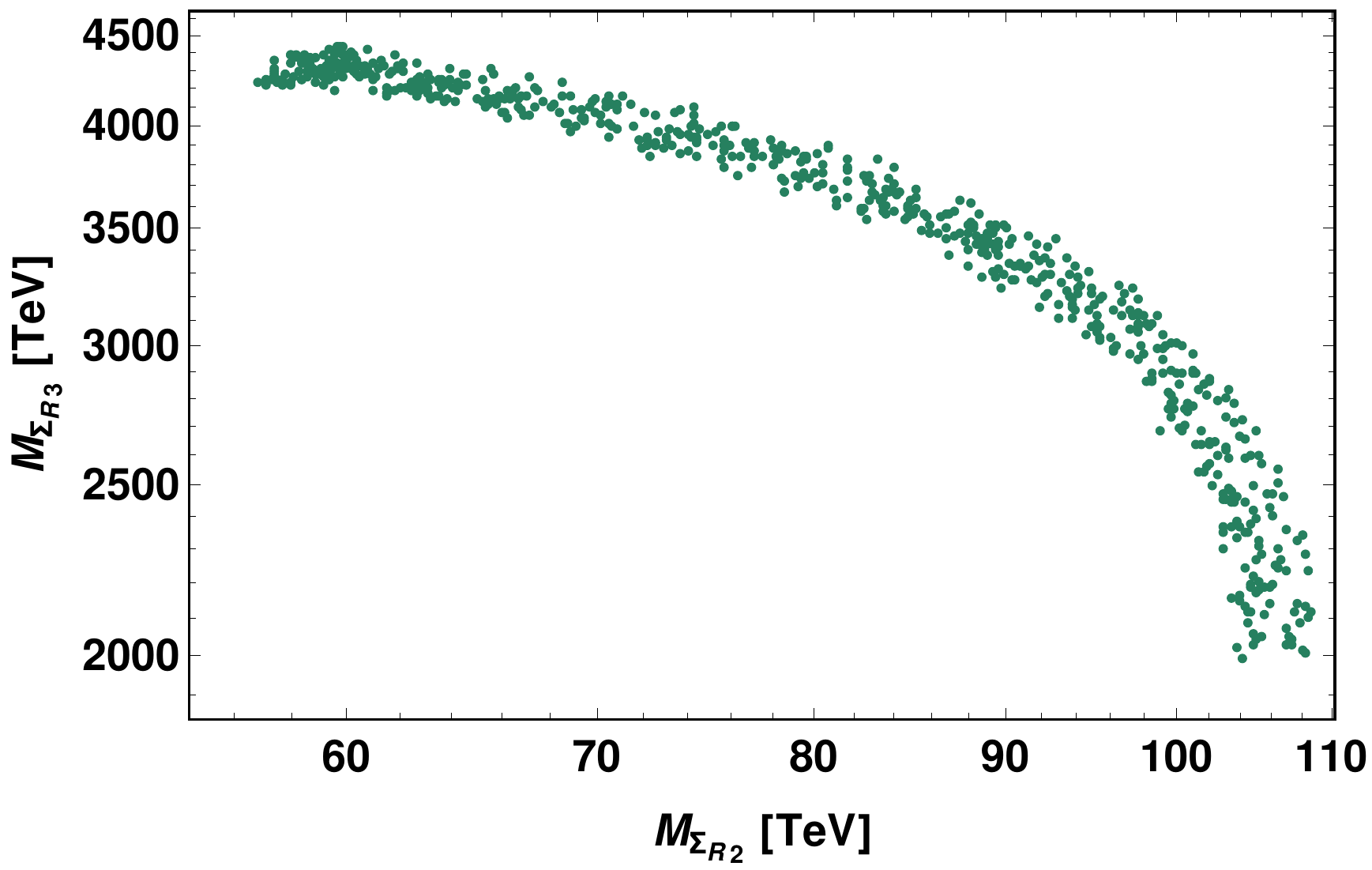}
\caption{The left (right) plots show the correlation between the heavy neutrino masses i.e., $M_{\Sigma_{R_1}}$ and $M_{\Sigma_{R_2}}$ ($M_{\Sigma_{R_2}}$ and $M_{\Sigma_{R_3}}$) in TeV scale.}
\label{M123}
\end{center}
\end{figure}

Fig. \ref{yukawa} shows the dependence of Yukawa couplings on the real and imaginary parts of $\tau$, while keeping the model parameters at their best-fit values. Finally, in Fig. \ref{M123} we show the hierarchical nature of the heavy neutrinos which follow the pattern $M_{\Sigma_{R_1}} \ll M_{\Sigma_{R_2}} \ll M_{\Sigma_{R_3}}$. 

\section{Leptogenesis}
\label{sec:lepto}
\begin{figure}[htpb]
\centering
\includegraphics[height=4.4cm, width=5.8cm]{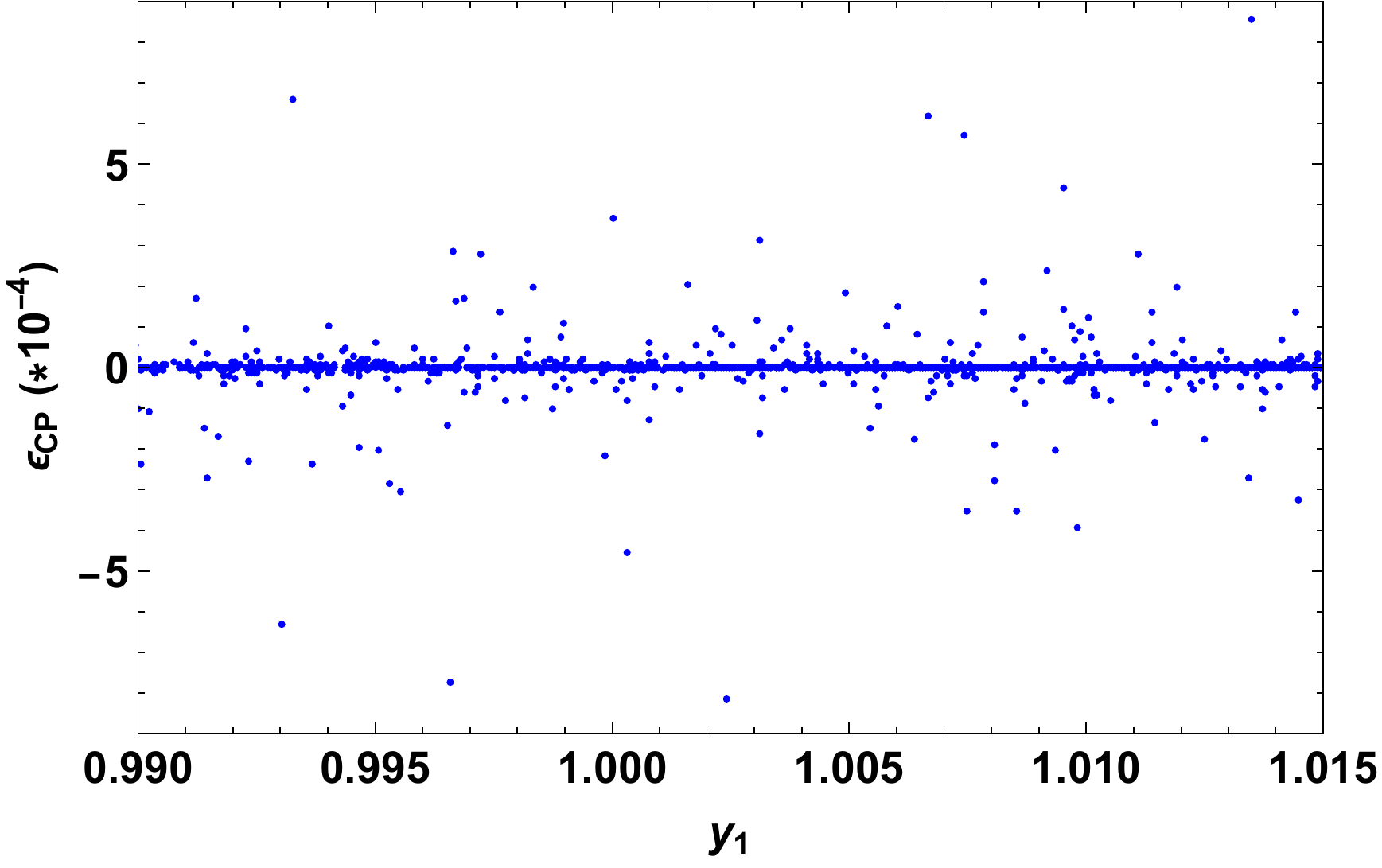}
\includegraphics[height=4.4cm, width=5.8cm]{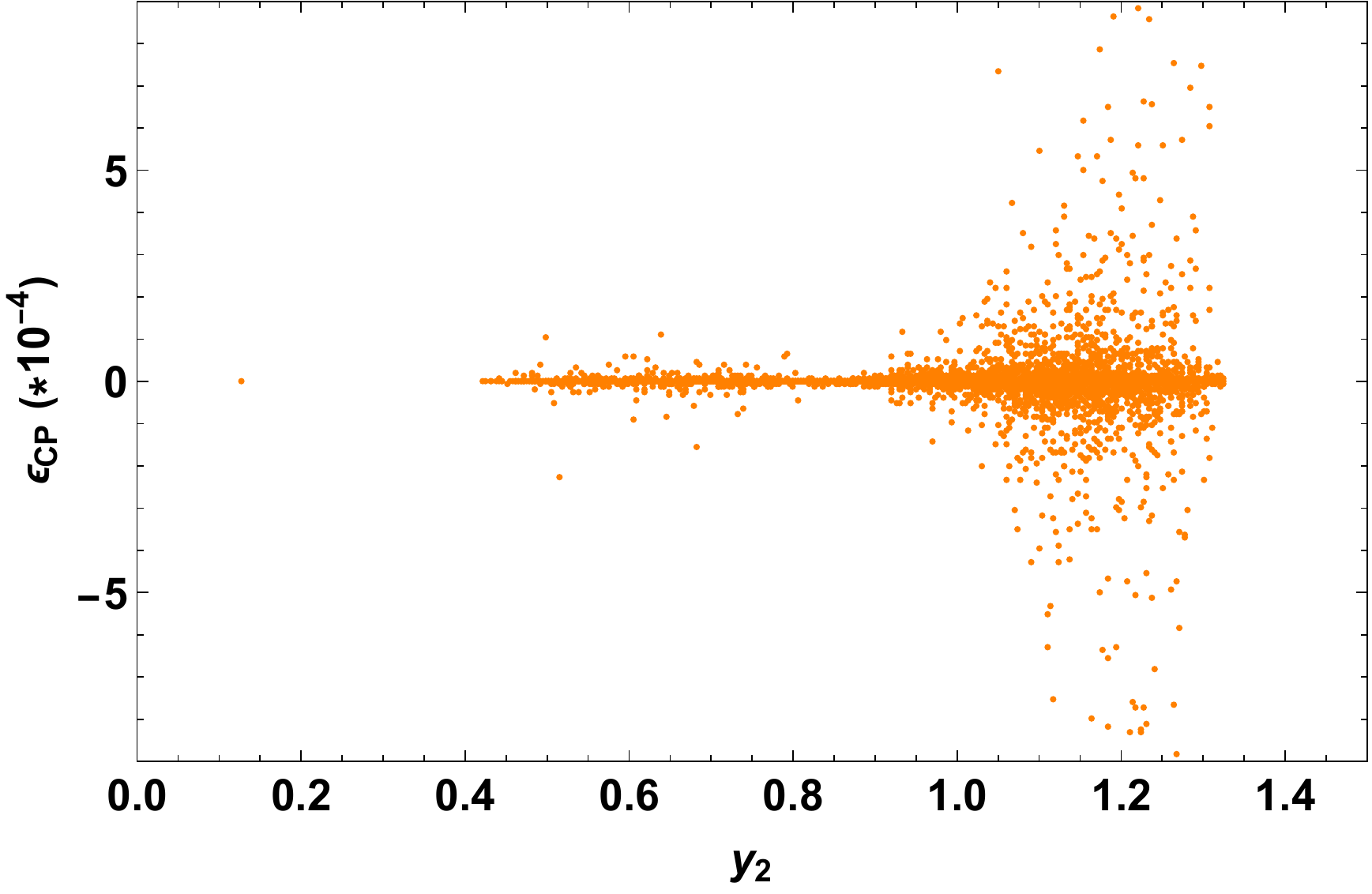}
\includegraphics[height=4.4cm, width=5.8cm]{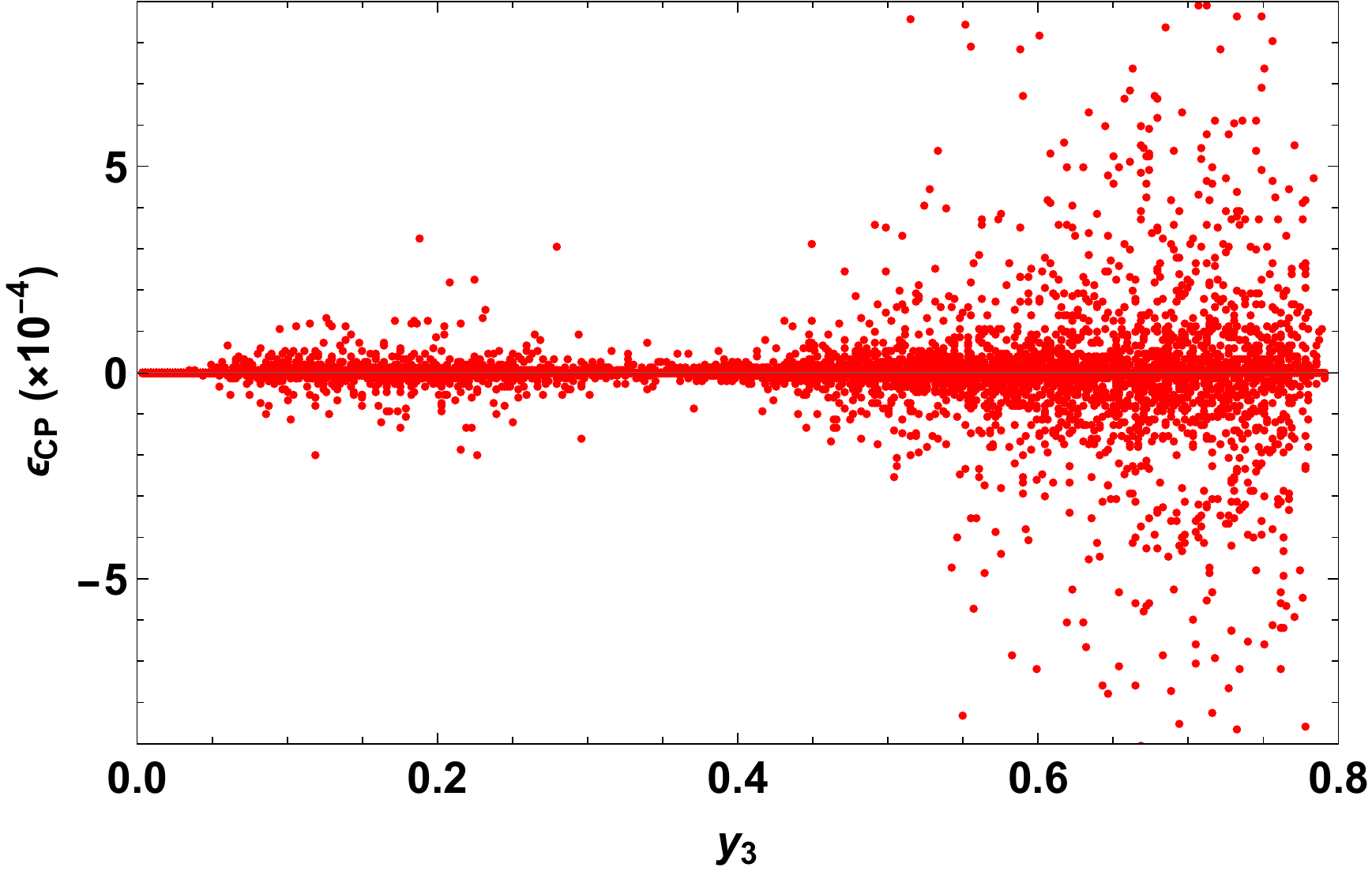}
\caption{In above we show the correlation of the Yukawa couplings i.e ($y_1$, $y_2$, $y_3$)} w.r.t CP asymmetry i.e., $\epsilon_{CP}$.
\label{CP_one_flav_y1y2y3}
\end{figure}

Considering the fact that the universe had started from an intially symmetric state of baryons and antibaryons, the present baryon asymmetry can be explained, as suggested by Sakharov \cite{Sakharov:1967dj},  if the following three criteria are satisfied:  Baryon number violation, C and CP violation and departure from thermal equilibrium during the evolution of the universe. Though the SM assures  all these criteria for an expanding universe akin ours, the extent of CP violation found in the SM is quite  small to accommodate the observed baryon asymmetry of the universe. Therefore, additional sources of CP violation are absolutely essential for explaining this asymmetry. The most common  new sources of CP violation possibly could arise in the lepton sector, which is however,   not yet  firmly established experimentally. Leptogenesis is the phenomenon that  furnishes a minimal setup to correlate the  CP violation in the lepton sector to the observed baryon asymmetry, as well as  imposes indirect constraints on the CP phases from the requirement that it would yield the correct baryon asymmetry. In here, we explore leptogenesis in type-III seesaw model with fermion triplets, where, the lightest heavy fermion is in TeV scale. The general expression for CP asymmetry is mentioned below \cite{Hambye:2012fh}
\begin{equation}
\epsilon_{CP}=-\sum_{j}\frac{3}{2} \frac{M_{\Sigma_{R_i}}}{M_{\Sigma_{R_j}}} \frac{\Gamma_{\Sigma_{R_i}}}{M_{\Sigma_{R_j}}} \left(\frac{V_j-2S_j}{3}\right) \frac{{\rm Im(\tilde Y_\Sigma \tilde Y_\Sigma^\dagger)^2}_{ij}}{{\rm(\tilde Y_\Sigma\tilde Y_\Sigma^\dagger)}_{ii} {\rm(\tilde Y_\Sigma \tilde Y_\Sigma^\dagger)}_{jj}},~~{\rm with}~~ \tilde{Y}_\Sigma= Y_\Sigma U_R,
\label{CP_asym}
\end{equation}
where, $Y_\Sigma=\left({M_D}/{v_u}\right)$ is the Yukawa matrix of Dirac mass term with its corresponding free parameters given in eqn. (\ref{eqn:Dirac}) and $U_R$ being the eigenvector matrix of $M_R$ used for its diagonalization i.e., $U_R M_R U_R^T\simeq  \mathrm{diag}  \{M_{\Sigma_{R_1}},M_{\Sigma_{R_2}},M_{\Sigma_{R_3}}\}$. From eqn.(\ref{CP_asym}) it is evident that vertex ($V_j$) and self-energy ($S_j$) diagrams \cite{Hambye:2012fh} must contribute to CP asymmetry significantly. However, in the hierarchical limit (i.e., $M_{\Sigma_{R_1}} \ll M_{\Sigma_{R_{2,3}}}$ and $M_{\Sigma_{R_2}} \neq M_{\Sigma_{R_3}}$) they attain the value unity i.e., ($S_j=V_j=1$). As, we don't have the hold on fine tuning of the Yukawa couplings, in order to calculate correct lepton asymmetry, we utilize the following benchmark values as shown in Table \ref{tab:leptobench}. Moreover, we also show in Fig. \ref{CP_one_flav_y1y2y3} the correlation between the one flavor CP asysmmetry i.e., $\epsilon_{CP}$ $\mathcal{O}(10^{-4})$\footnote{It is to note that the ranges of the Yukawa couplings are same as in neutrino sector but in Fig. \ref{yukawa} the plots are expressed in log scale while implementing $\chi^2$ minimisation, hence, suppressing the upper bounds and magnifying the lower bounds more prominently, therefore the ranges might look different due to different scales utilized.} with the Yukawa couplings within their corresponding
ranges i.e., $0.99 \lesssim y_1 \lesssim 1.015$  (upper left panel), $0.4 \lesssim y_2 \lesssim 1.3$ (upper right panel) and $0.1 \lesssim y_3 \lesssim 0.8$ (bottom panel).

\begin{table}[htpb]
\begin{center}
\centering
\begin{tabular}{|c|c|c|c|c|c|}  \hline
~$y_1$~  & ~$y_2$~& $y_3$ & { $M_{\Sigma_{R_1}}$} &  {$\epsilon_{CP}$}\\\hline 
~$1.0051$ ~ &  ~  $0.574$~ & ~ $0.312$~&~ $6.53 \times 10^{3}$ $\rm GeV$~& ~ $9.713\times 10^{-4}$~ \\\hline
\end{tabular}
\caption{Benchmark values of the Yukawa couplings and CP asymmetry utilized to generate the correct lepton asymmetry.}
\label{tab:leptobench}
\end{center}
\end{table}

\subsection{Boltzmann Equations} 
The dynamics of applicable Boltzmann equations determine the evolution of particle number densities. The Sakharov conditions \cite{Sakharov:1967dj} necessitate the decay of the parent heavy fermion, which must be out of equilibrium in order to generate the lepton asymmetry. To do so, one must compare the Hubble expansion rate to the decay rate, as shown below:
\begin{equation}
K_{\Sigma_{R_i}} = \frac{\Gamma_{\Sigma_{R_i}}}{H(T=M_{\Sigma_{R_i}})}.
\end{equation}
The Hubble rate is defined as $H = \frac{1.67 \sqrt{g_\star}~ T^2 }{M_{\rm Pl}}$, where, $g_{\star} = 106.75$ is the number of relativistic degrees  of  freedom in the thermal bath and $M_{\rm Pl} = 1.22 \times 10^{19}$ GeV is the Planck mass. The size of the couplings between the triplet fermions and leptons become the determining factor, guaranteeing that inverse decay does not approach thermal equilibrium. For example, if the value is less than or equal to $10^{-7}$, it gives $K_{\Sigma_{R_i}} \sim 1$. The Boltzmann equations associated with evolution of the number densities of right-handed fermion field and lepton can be articulated in terms of the yield parameters, i.e.,  the ratio of number densities to entropy density, and are expressed as \cite{Plumacher:1996kc,Giudice:2003jh,Buchmuller:2004nz,Iso:2010mv}
\begin{eqnarray}
&& \frac{d Y_{\Sigma}}{dz}=-\frac{z}{s H(M_{\Sigma})} \left[\left( \frac{Y_{\Sigma}}{{Y^{\rm eq}_{\Sigma}}}-1\right)\gamma_D +\left( \left(\frac{{Y_{\Sigma}}}{{Y^{\rm eq}_{\Sigma}}}\right)^2-1\right)\gamma_A \right],\nonumber\\
&& \frac{d Y_{{B-L}}}{d z}= -\frac{z}{s H(M_{\Sigma})} \left[ \frac{Y_{{B-L}}}{{Y^{\rm eq}_{\ell}}}\gamma_{D}- \epsilon_{CP} \left( \frac{Y_{\Sigma}}{{Y^{\rm eq}_{\Sigma}}}-1\right)\frac{\gamma_D}{2} \right],\label{Boltz1}
\end{eqnarray}
where $z = M_{\Sigma_{R_i}}/T$, $s$ is the entropy density,   and the equilibrium number densities have the form \cite{Davidson:2008bu}
\begin{eqnarray}
Y^{\rm eq}_{\Sigma}= \frac{135  g_{\Sigma}}{16 {\pi}^4 g_\star} z^2 K_2(z), \hspace{3mm} {Y^{\rm eq}_\ell}= \frac{3}{4} \frac{45 \zeta(3) g_\ell}{2 {\pi}^4 g_{\star}}\,.\label{eq:39}
\end{eqnarray}
 $K_{1,2}$ in Eq. (\ref{eq:39})  represent the modified Bessel functions,  the lepton and RH fermion degrees of freedom take the values $g_\ell=2$ and $g_{\Sigma_{R_i}}=2$ and the decay rate $\gamma_D$ is given as
\begin{eqnarray}
\gamma_D &=& s Y^{\rm eq}_{\Sigma}\Gamma_{\Sigma} \frac{K_1(z)}{K_2(z)},~~~ \Gamma_\Sigma= \frac{1}{8\pi} M_{\Sigma_{R_i}}(\tilde{\rm{Y}}_\Sigma^\dagger\tilde{\rm{Y}}_\Sigma)_{ii}, \nn\\
\gamma_A &=& \frac{M_{\Sigma_{R_1}} T^3}{32 \pi^3}e^{-2z}\left[\frac{111g^4}{8\pi}+\frac{3}{2z} \left(\frac{111g^4}{8\pi} +\frac{51g^4}{16\pi}\right)+\mathcal{O}(1/z)^2\right],
\end{eqnarray}
wherein $\gamma_A$ denotes the gauge annihilation process \cite{Hambye:2012fh,Mishra:2020gxg}, with $g$ being the typical gauge coupling. 
\begin{figure}[htpb]
\begin{center}
\includegraphics[height=50mm,width=75mm]{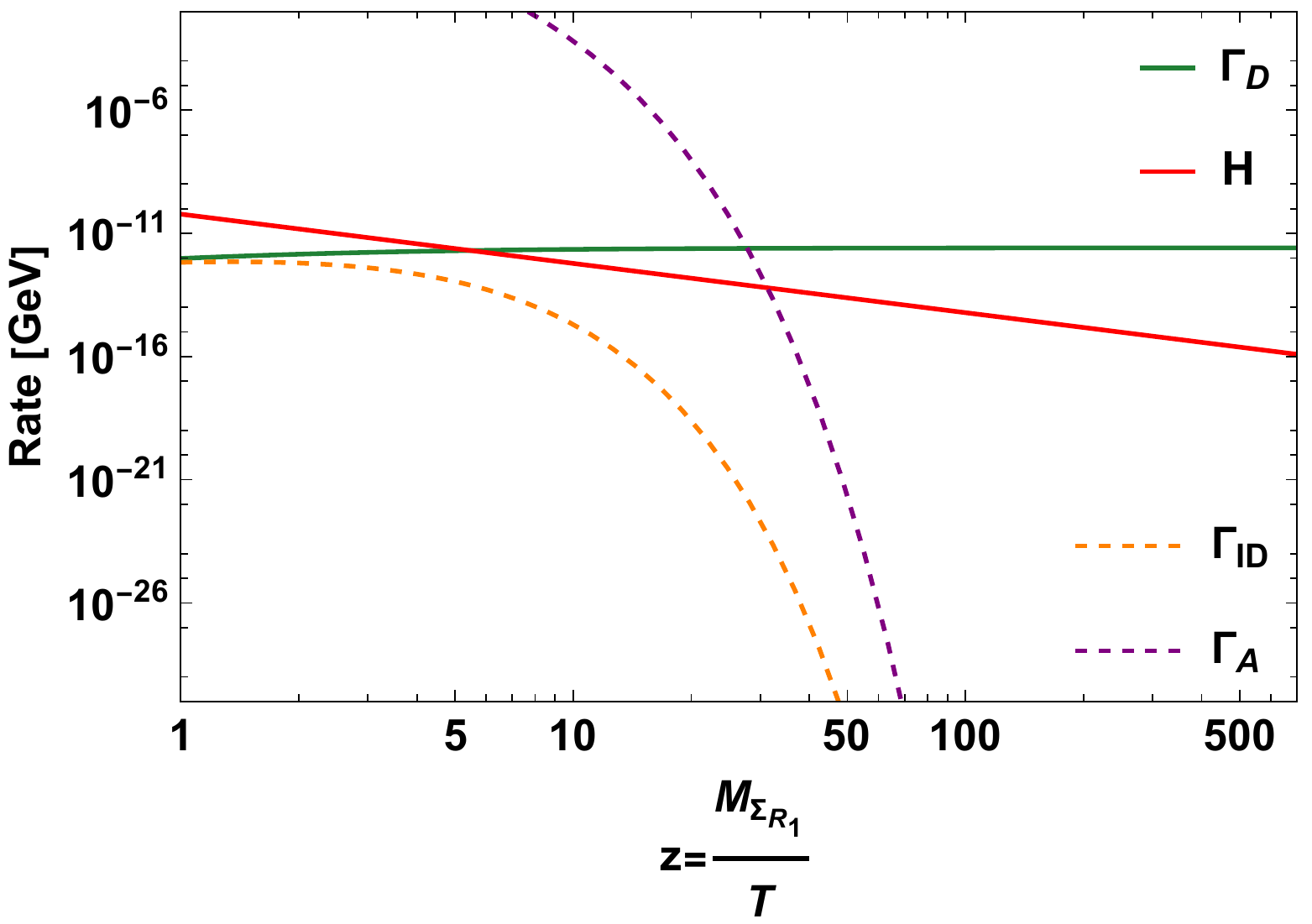}
\hspace{2mm}
\includegraphics[height=50mm,width=75mm]{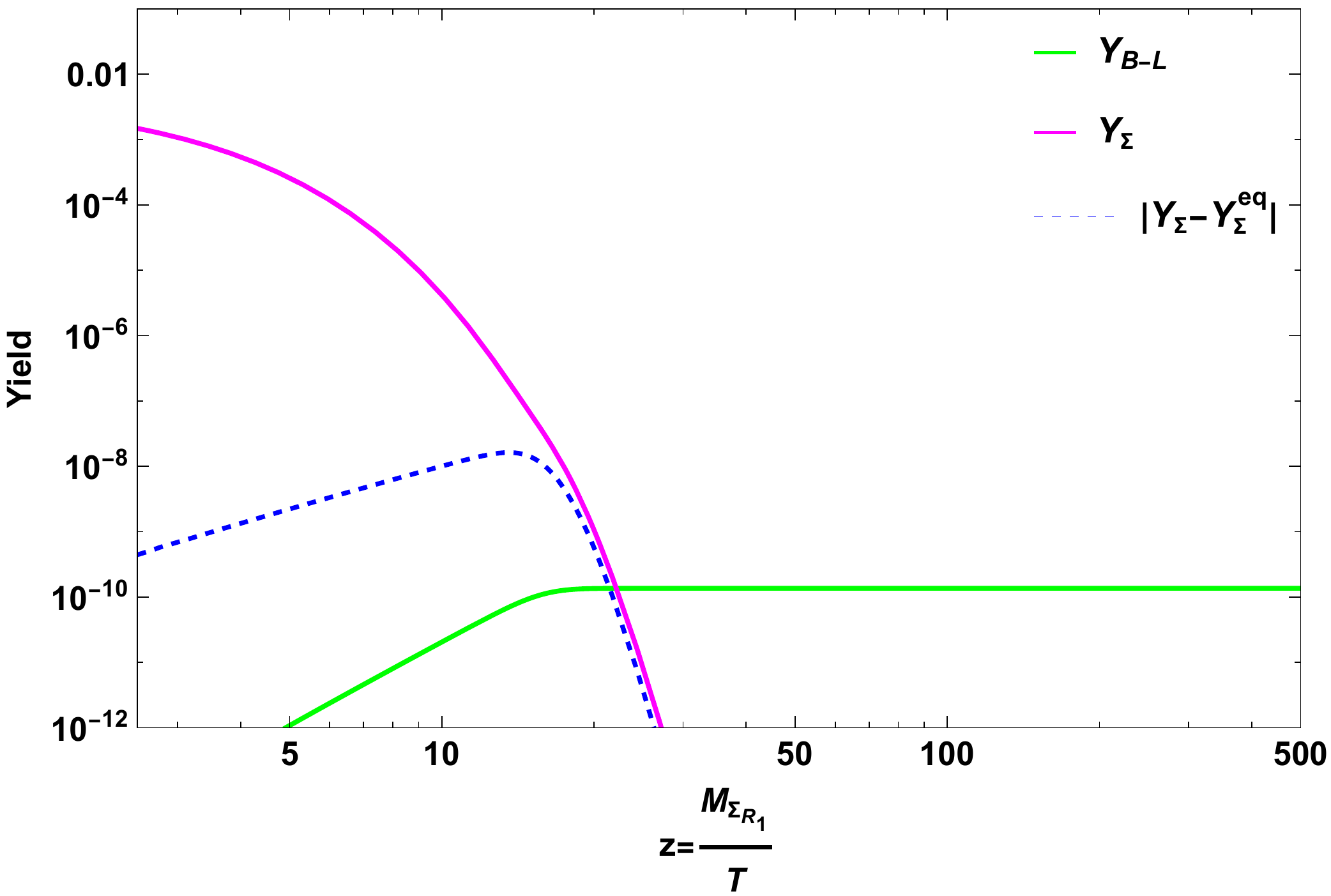}
\caption{Left panel exhibits the comparison of interaction rates with Hubble expansion rate ($H$) represented by red solid line, where, green solid line corresponds to decay $\left(\Gamma_D = \Gamma_\Sigma \frac{K_1(z)}{K_2(z)} \right)$ \cite{Davidson:2012fn}, inverse decay ($\Gamma_{ID} = \Gamma_{D} \left(\nicefrac{Y_\Sigma^{eq}}{Y_\ell^{eq}}\right)$) \cite{Strumia:2006qk} is shown by dotted orange line and annihilation rate ($\Gamma_A$) by dotted purple line. Right panel projects the evolution of $Y_{B-L}$ (green solid line) as a function of $z = \frac{M_{\Sigma_{R_1}}}{T}$.}
\label{yield}
\end{center}
\end{figure}
The comparison of the interaction rates with Hubble expansion rate ($H$) is displayed in the left panel of Fig. \ref{yield}, while the solution of Boltzmann eq. (\ref{Boltz1}) is presented in the right panel. For coupling strength of around ($\simeq~ 10^{-7}$), $Y_{\Sigma}$ (magenta solid curve) with $|Y_\Sigma-Y^{\rm eq}_{\Sigma}|$ (blue dashed curve) are shown where the generated lepton asymmetry is around ($\simeq10^{-10}$) (green thick curve). The lepton asymmetry thus obtained can be converted into baryon asymmetry through the sphaleron transition process, and is  given as \cite{Plumacher:1996kc,Vatsyayan:2022rth,Harvey:1990qw} 
\begin{equation}
Y_B = 3\left(\frac{8n_f + 4 n_H}{22 n_f + 13 n_H}\right)Y_{{B-L}},
\end{equation}
where, $n_f$ represents the number of triplet fermion generations,  $n_H$ denotes the no. of Higgs doublets and the factor of 3 comes from the three $SU(2)_L$ degrees of freedom of
the triplets. The observed baryon asymmetry of the universe generally expressed  in terms of baryon to photon ratio as \cite{Planck:2018vyg}
\begin{equation}
\eta = \frac{n_b - n_{\bar{b}}}{n_\gamma} = 6.08 \times 10^{-10}.
\end{equation}
The current bound on baryon asymmetry {\cite{AharonyShapira:2021ize}} can be procured from the relation $Y_{B} = \eta/ 7.04$ as {$Y_{B} = (8.6 \pm 0.1)\times 10^{-11} \equiv Y^{obs}_B$}. Using the asymptotic value  of the lepton asymmetry as ($8.77 \times 10^{-10}$)  from Fig. \ref{yield}, we obtain the value of baryon asymmetry as {$Y_{B} = \frac{24}{23}~Y_{B-L} \sim 10^{-10}$}.

\subsection{A note on flavor consideration}

When ($T>10^{12}$ GeV), one flavor approximation suffices in leptogenesis, indicating that all Yukawa interactions are out of equilibrium. However, at temperatures $\ll 10^8$ GeV, various charged lepton Yukawa couplings (i.e., each for three generations) come into equilibrium, making flavor effects a crucial factor in determining the final lepton asymmetry. All Yukawa interactions occur in equilibrium at temperatures below $10^6$ GeV, and the asymmetry is encoded in the individual lepton flavor. Numerous studies on flavor effects in type-I leptogenesis can be found in the literature \cite{Pascoli:2006ci,Antusch:2006cw,Nardi:2006fx,Abada:2006ea,Granelli:2020ysj,Dev:2017trv}. The lower bounds on heavy Majorana masses are relaxed when flavour effects are taken into account, giving more room for low scale leptogenesis \cite{Abada:2018oly,Drewes:2022kap,Alanne:2018brf}. Given the significance of flavour effects in low scale leptogenesis, we briefly examine their implications in the current framework in relation to the CP asymmetry for each particular lepton flavour ($\alpha=e,\mu,\tau$) given below \cite{Dev:2015cxa,Mishra:2019gsr} 
\begin{figure}[htpb]
\centering
\includegraphics[height=5cm,width=8.5cm]{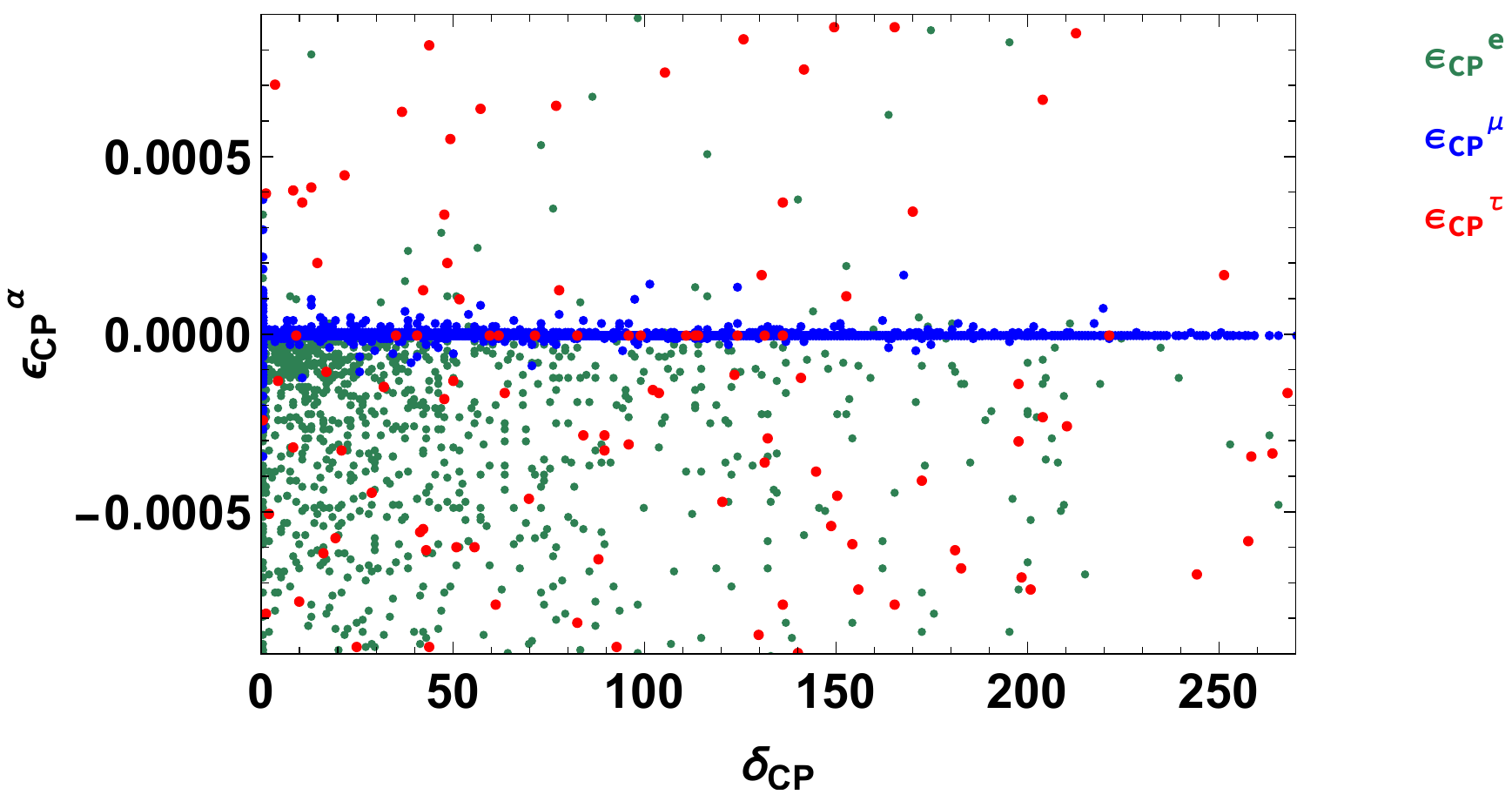}
\caption{In the above plot the interdependence of CP asymmetry corresponding to each flavor i.e., ($\alpha=e,\mu,\tau$) is been projected against $\delta_{CP}$.}
\label{fig:FCP_asym}
\end{figure}

\begin{equation}
\epsilon^{\alpha}_{\Sigma}=\frac{1}{2}\sum_{j} \frac{M_{\Sigma_{R_i}}}{M_{\Sigma_{R_j}}} \frac{\Gamma_{\Sigma_{R_i}}}{M_{\Sigma_{R_j}}}  \left[\frac{\rm{Im}\left[ \left(\tilde Y_\Sigma \tilde Y_\Sigma^\dagger\right)_{ij} \tilde Y_{\alpha i}^* \tilde Y_{\alpha j}\right]}{{\rm{(\tilde Y_\Sigma\tilde Y_\Sigma^\dagger)}}_{ii} {\rm(\tilde Y_\Sigma \tilde Y_\Sigma^\dagger)}_{jj}}\right],
\label{CP_asym}
\end{equation}
\begin{figure}[htpb]
\begin{center}
\includegraphics[height=58mm,width=75mm]{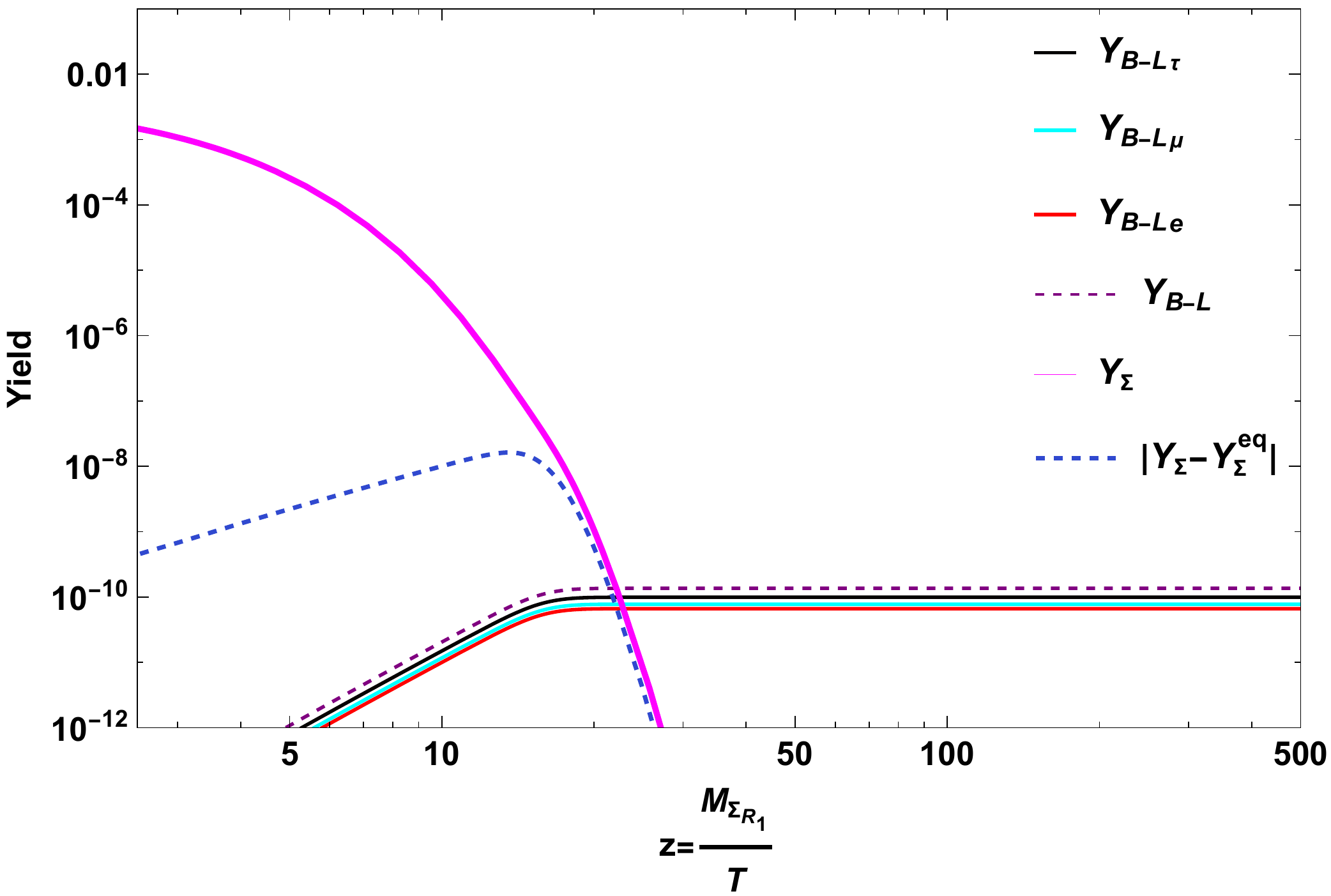}
\vspace{7mm}
\includegraphics[height=58mm,width=75mm]{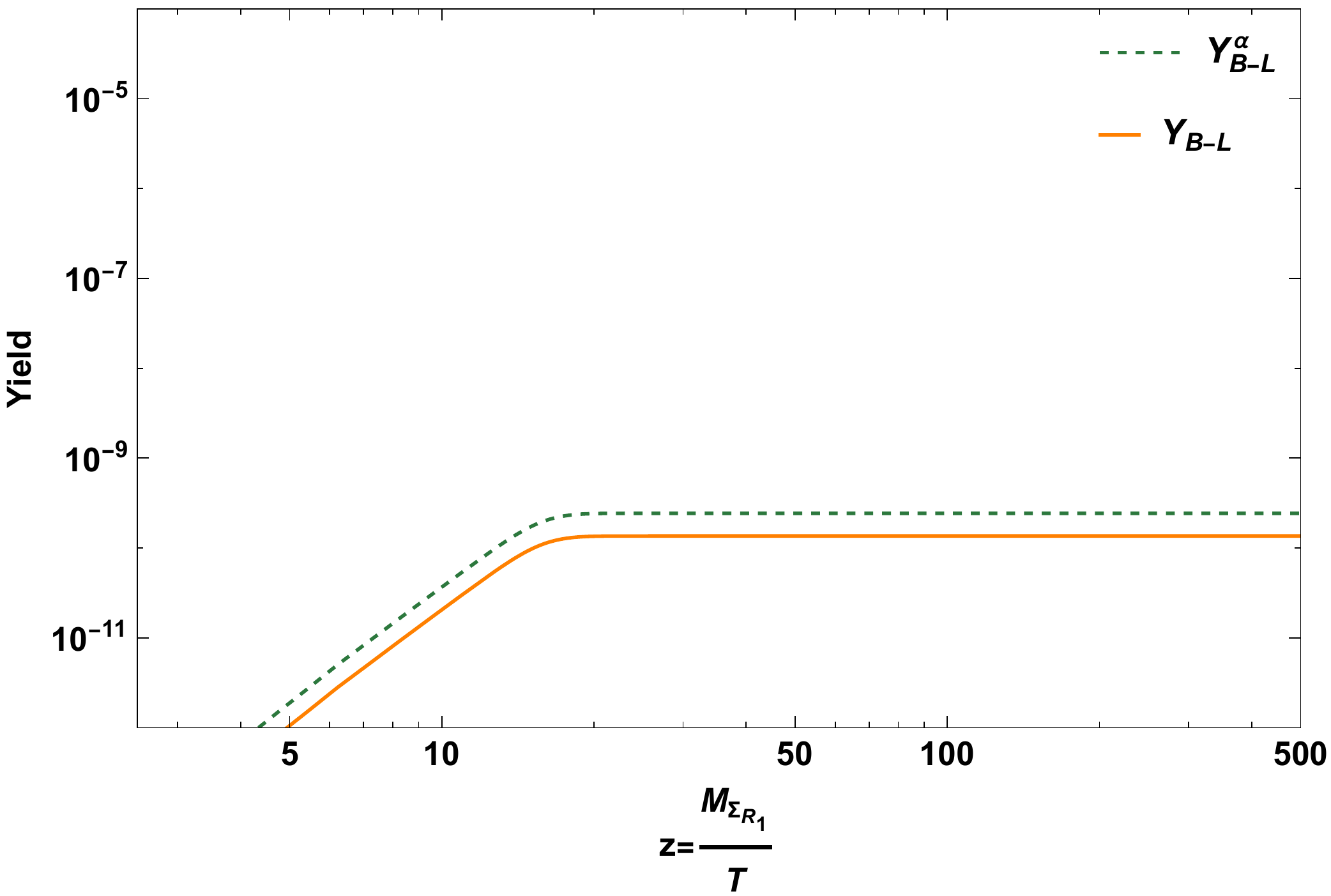}
\caption{After including the flavor effects the yield is shown in left panel, whereas, right panel displays the yield enhancement  due to flavor effects.}
\label{yield_Flavor}
\end{center}
\end{figure}
The Boltzmann equation describing the generation of  $(B-L)$ asymmetry for each lepton  flavor is \cite{Antusch:2006cw}
\begin{eqnarray}
\frac{d Y^{\alpha}_{ B-L}}{d z}= -\frac{z}{s H(M_\Sigma)} \left[  \epsilon^\alpha_{\Sigma} \left( \frac{Y_{\Sigma}}{{Y^{eq}_{\Sigma}}}-1\right)\gamma_D-\left(\frac{\gamma^{\alpha}_D}{2}\right)\frac{A_{\alpha \alpha}Y^\alpha_{\rm B-L_\alpha}}{{Y^{eq}_{\ell}}}\right],
\end{eqnarray}
where, $\epsilon^\alpha_{\Sigma}$ i.e., $(\alpha=e,\mu,\tau)$ represents the CP asymmetry in each lepton flavor
\begin{equation}
\gamma_D^\alpha = s Y_{\Sigma}^{eq}\Gamma_{\Sigma}^\alpha \frac{K_1(z)}{K_2(z)}, \quad \gamma_D = \sum_\alpha \gamma^\alpha_D.\nn\\
\end{equation}
The matrix {$A$} is given by \cite{Nardi:2006fx}, 
\begin{equation}
{A}=\begin{pmatrix}
-\frac{221}{711} && \frac{16}{711} && \frac{16}{711}\\
\frac{16}{711} && -\frac{221}{711} && \frac{16}{711}\\
\frac{16}{711} && \frac{16}{711}  && -\frac{221}{711} \\
\end{pmatrix}.\nn\\
\end{equation}
In addition to which we have expressed a plot to show the interdependence of each flavor on $\delta_{CP}$ in Fig. \ref{fig:FCP_asym}. Subsequently, for the flavor case, benchmark values of CP asymmetry associated with ($e,\mu, \tau$) flavors are $\epsilon_{\Sigma}^e =4.7\times10^{-4}$, $\epsilon_{\Sigma}^\mu =5.6\times 10^{-4}$ and $\epsilon_{\Sigma}^\tau =7.2\times10^{-4}$ respectively.  Therefore, we estimate the $B-L$ yield with flavor consideration in the left panel of Fig. \ref{yield_Flavor}. It is quite obvious to notice that the enhancement in $B-L$ asymmetry is obtained in case of flavor consideration (green dashed line) over the one flavor approximation (orange solid line), as displayed in the right panel. This is because, in one flavor approximation the decay of the heavy fermion to a particular lepton flavor final state can get washed away by the inverse decays of any flavor unlike the flavored case \cite{Abada:2006ea}.
\section{Collider Bound on  $Z'$ mass}
\label{sec:Zmass}
As previously mentioned in Sec \ref{sec:2}, the $U(1)_{B-L}$ gauge symmetry is spontaneously broken by assigning the vacuum expectation value $v_\rho$ to the singlet scalar $\rho$. Consequently, the   neutral gauge boson $Z'$ associated with this symmetry  becomes massive by absorbing the massless pseudoscalar component of $\rho$ and its mass is given as,
\begin{eqnarray}
M_{Z'}= g_{BL} v_\rho\;,
\end{eqnarray}
 where, $g_{BL}$ is the gauge coupling constant of $U(1)_{B-L}$. The LEP-II provides  the constraint on the ratio of mass of $Z'$ boson to its coupling as
$M_{Z'}/g_{BL} >6.9$  TeV \cite{ALEPH:2013dgf}. Hence, in this work we have considered the range of the   $v_\rho$ as [$10^3-10^4$] TeV (\ref{ranges}), consistent with the LEP-II bound.

The ATLAS and CMS collaborations have performed extensive searches for the new resonances in both dilepton and dijet channels. In the absence of any excess events over the SM background, they put lower bounds on the mass of $Z'$ boson. These bounds are usually limited to a specific model, and typically the experiments report their results assuming simplified models, like the Sequential Standard Model (SSM) or GUT-inspired $E_6$ models.

\begin{figure}
\begin{center}
\includegraphics[height=48mm,width=75mm]{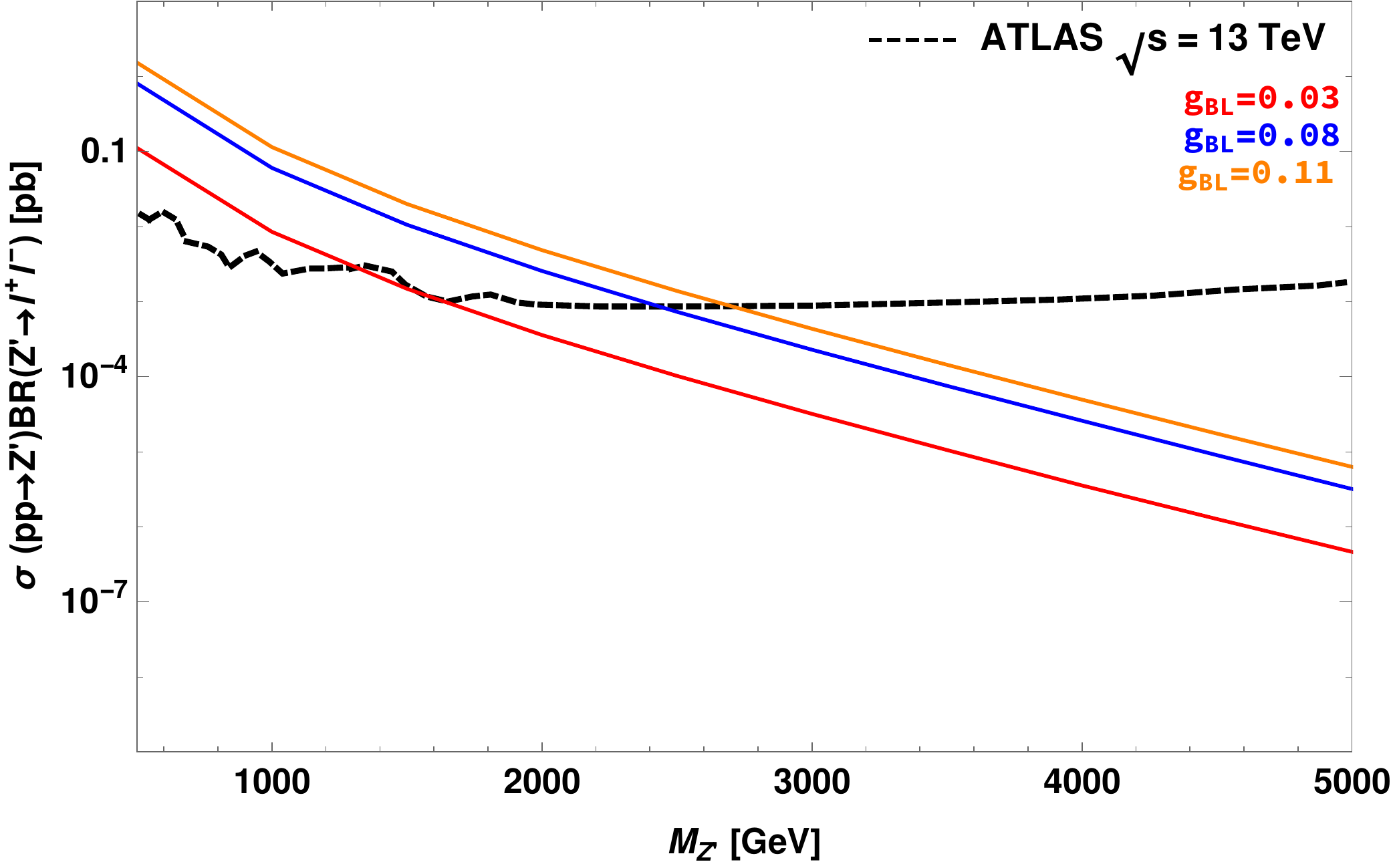}
\caption{ {The colored lines represent the dilepton signal cross sections for $pp \to Z' \to ee(\mu \mu)$ as a function of $M_{Z'}$ for a representative set of $g_{BL}$ values  and the black dashed line  symbolizes the ATLAS bound \cite{ATLAS:2019erb}.}}\label{Zpmass}
\end{center}
\end{figure}
  
Recent results from ATLAS \cite{ATLAS:2019erb}, provide the lower limits on the $Z'$ mass  from the dilepton search using Run 2 data, collected with the center of mass energy   $\sqrt s =13$ TeV. In  this  work,  we  use  CalcHEP  \cite{Belyaev:2012qa} to  compute  the  production  cross section of $Z'$, i.e., $pp \to Z' \to ee(\mu \mu)$.  In  Fig. \ref{Zpmass},  we show the $Z'$ production cross section times the branching fraction of $Z'$ decaying to dilepton ($ee,\mu \mu$) signal as a function of $M_{Z'}$, for some representative values of the gauge coupling $g_{BL}=0.03,0.08,0.11$.  The black dashed line denotes the dilepton bound from ATLAS \cite{ATLAS:2019erb}.  It can be noticed from the figure  that the region below $M_{Z'}\simeq $ 1.3 TeV  is  excluded  for $g_{BL}=  0.03$ {in red color}.   For $g_{BL}=  0.08$, $M_{Z'}< 2.47$  TeV {in blue color}  is  ruled  out   and  the  mass  region  of $M_{Z'} > 2.69$  TeV  is  allowed  for $g_{BL}= 0.11$ {in orange color}.  Thus, one can generalize these observations as the lower limits on $M_{Z'}$ increases with the increase of the gauge couplings.
\section{Conclusion}
\label{sec:conclude}
We have curated a model involving $ \rm A_4$ modular symmetry  and $U(1)_{B-L }$ gauged symmetry using type-III seesaw mechanism in super-symmetric context in order to realize the neutrino phenomenology and to explain the observed oscillation data.  We have incorporated  $SU(2)_L$ triplet fermions ($\Sigma$) along with a singlet weighton field ($\rho$). The Yukawa couplings acquire modular forms under $A_4$ modular discrete symmetry, where, acquisition of VEV by modulus $\tau$ breaks $A_4$ symmetry.  This discrete symmetry is useful in procuring a definite neutrino mass matrix structure. Here, in analysis section numerical diagonalization technique lessens the burden and the predicted results are in accordance to the 3$\sigma$ bound as obtained through several experiments. We can extract the best fit values for the model parameters using the chi-square minimization approach, which helps us find strong correlations between the observables.  As a consequence, we obtain the sum of active neutrino masses $\sum m_{\nu_i}$ within $[0.058-0.12]$ eV and mixing angles are seen to be within their 3$\sigma$ ranges. The model engenders neutrinoless double beta decay mass parameter $\langle m_{ee} \rangle $ between 0.0039 and 0.0087, which assures the limit coming from KamLAND-Zen experiment. Also, Majorana phases $\alpha_{21}$ and $\alpha_{31}$ are revealed in the range $[0^\circ,80^\circ]$ and $[0^\circ,360^\circ]$ respectively. Proceeding further,   the results for  $\delta_{\rm CP}$ and Jarlskog invariant $J_{\rm CP}$ is seen to be within $[202^{\circ},211^\circ]$ and [5.5,7.1] $\times 10^{-3}$ respectively establishing a strong correlation.
Further, as there is an hierarchical mass difference between the heavy fermions ($M_\Sigma$) in the model with $M_{\Sigma_{R_1}}$, $M_{\Sigma_{R_2}}$ and $M_{\Sigma_{R_3}}$ are found to be  within the range [$6-8.5$] TeV, [$50-110$] TeV and [$2000-4500$] TeV respectively, hence, the decay of lightest one gives rise to non-zero CP asymmetry. The lepton asymmetry coming from Boltzmann equation is $\simeq$ $10^{-10}$, and hence explains the baryon asymmetry of the Universe and also we have discussed the flavor effects as our lightest heavy fermion is in TeV scale. Additionally, we have discussed the mass of the new neutral $Z^\prime$ gauge boson associated with $U (1)_{B-L}$ symmetry which is within the present experimental collider bounds.
\section*{Acknowledgements}
PM and PP want to thank Prime Minister’s Research Fellowship (PMRF) scheme for its
financial support. MKB wants to thank DST-Inspire for financial help. RM would like to
acknowledge University of Hyderabad IoE project grant no. RC1-20-012. The use of CMSD HPC facility of Univ. of Hyderabad to carry out computational work is duly acknowledged. We thank Purushottam Sahu and Dr. Shivaramakrishna Singirala for useful discussion.
\appendix
\section{$A_4$ modular symmetry}
\label{Appendix}
\
$ \rm A_4$ group is the alternating group of even permutations of four entries. 
 It is isomorphic to the tetrahedral symmetry. 
The generators of the group S and T, following the relations,
\begin{equation}
S^2 = (TS)^3 = (ST)^3 = \mathbb{I}.
\end{equation} \\
Group formed by the generators S and T is the inhomogeneous modular group $\bar{\Gamma}$ and the transformations are abbreviated as follows \citep{King:2020qaj,feruglio2019neutrino}\\
\begin{equation}
S: \tau \rightarrow \frac{-1}{\tau}, \hspace{2cm} T : \tau \rightarrow \tau +1
\end{equation}Representation of S and T in the SL(2,$\mathbb{Z}$ ) group is,\\
\begin{equation}
S=
\begin{pmatrix}
0 & 1 \\
-1 & 0
\end{pmatrix}, \hspace{2cm}
T = 
\begin{pmatrix}
1 & 1 \\
0 & 1
\end{pmatrix}.
\end{equation} 
A group of linear fractional transformations forms the modular group, which transforms the modulus $\tau$  in the upper half-plane [Im($\tau$) $>$ 0],

\begin{equation}
\tau \rightarrow \frac{c \tau + d}{a \tau + b} ,\hspace{2cm} (a, b, c, d \hspace{1mm} are\hspace{1mm} integers,\hspace{1cm}  cb-da=1)
\end{equation}\\
and the mapping,
\begin{equation}
\frac{c \tau + d}{a \tau + b} \rightarrow  
\begin{pmatrix}
c & d \\
a & b
\end{pmatrix},
\end{equation}\\
is an isomorphism from the modular group. Following is the series of groups $\Gamma(N)$, where N=1, 2, 3...,\\
\begin{equation}
 \Gamma(N)= \Big\{
 \begin{pmatrix}
 c & d \\
a & b
 \end{pmatrix} \in SL(2,\mathbb{Z}), \hspace{2mm}
 \begin{pmatrix}
 c & d \\
a & b
 \end{pmatrix} =
  \begin{pmatrix}
 1 & 0 \\
 0 & 1 
 \end{pmatrix} \Big\},
\end{equation} \\
where, $\Gamma$ = SL(2,$\mathbb{Z}$ ) is homogeneous modular group and $\Gamma$(N). The group $\Gamma$(N) operates on the complex modulus $\tau$, in the upper half plane as the linear fractional transformation,
\begin{equation}
\gamma = \frac{c \tau + d}{a \tau + b}.
\label{gamma}
\end{equation}\\
A significant modular invariant element is the modular function $f(\tau)$ which is holomorphic function of $\tau$ with level N and modular weight 2k, under  $\Gamma(N)$ is,
\begin{equation}
f\Big(\frac{c \tau + d}{a \tau + b}\Big) = (a\tau + b)^{2k}f(\tau), \hspace{2mm}\forall
\begin{pmatrix}
c & d \\
a & b
\end{pmatrix} = \gamma \in {\Gamma}(N).
\end{equation}
Here, N can vary according to the symmetry group $\rm A_4$, $\rm S_3$, $\rm S_4$, or $\rm A_5$. In reference \citep{King:2020qaj},  it is given that for N = 2, 3, 4, and 5; $\Gamma_2$, $\Gamma_3$, $\Gamma_4$ and $\Gamma_5$ are isomorphic to $\rm S_3$, $\rm A_4$, $\rm S_4$, and $\rm A_5$ respectively. 
The modular forms of $A_4$ triplet Yukawa couplings read as,
\begin{eqnarray}
Y_{1}(\tau)&=& \frac{i}{2\pi}\left[\frac{\eta^{\prime}(\frac{\tau}{3})}{\eta(\frac{\tau}{3})} + \frac{\eta^{\prime}(\frac{\tau+1}{3})}{\eta(\frac{\tau+1}{3})} +\frac{\eta^{\prime}(\frac{\tau+2}{3})}{\eta(\frac{\tau+2}{3})} -\frac{27\eta^{\prime}(3\tau)}{\eta({3\tau})} \right] \label{mod_y1}\\
Y_{2}(\tau)&=& \frac{-i}{\pi}\left[\frac{\eta^{\prime}(\frac{\tau}{3})}{\eta(\frac{\tau}{3})} +\omega^2 \frac{\eta^{\prime}(\frac{\tau+1}{3})}{\eta(\frac{\tau+1}{3})} +\omega\frac{\eta^{\prime}(\frac{\tau+2}{3})}{\eta(\frac{\tau+2}{3})}  \right] \label{mod_y2}\\
Y_{2}(\tau)&=& \frac{-i}{\pi}\left[\frac{\eta^{\prime}(\frac{\tau}{3})}{\eta(\frac{\tau}{3})} +\omega \frac{\eta^{\prime}(\frac{\tau+1}{3})}{\eta(\frac{\tau+1}{3})} +\omega^2\frac{\eta^{\prime}(\frac{\tau+2}{3})}{\eta(\frac{\tau+2}{3})}  \right]
\label{mod_y3}
\end{eqnarray}
 where $\eta(\tau)$ is Dedekind eta-function which can be defined in the upper half plane of the complex plane.
 \begin{align}
 \eta(\tau)= q^{1/24} \prod_{m=1}^{\infty} (1-q^m)~~~~~~~~ q = e^{\iota2\pi\tau}
 \label{dedekind}
 \end{align}

\bibliographystyle{my-JHEP}
\bibliography{type3}

\end{document}